# Photonic frequency microcombs based on dissipative Kerr and quadratic cavity solitons

MINGMING NIE,[1, *] YIJUN XIE,[1] BOWEN LI,[1] AND SHU-WEI HUANG[1, *]

[1]*Department of Electrical, Computer and Energy Engineering, University of Colorado Boulder, Boulder, Colorado 80309, USA*
**mingming.nie@colorado.edu*, *shuwei.huang@colorado.edu*

**Abstract:** Optical frequency comb, with precisely controlled spectral lines spanning a broad range, has been the key enabling technology for many scientific breakthroughs. In addition to the traditional implementation based on mode-locked lasers, photonic frequency microcombs based on dissipative Kerr and quadratic cavity solitons in high-Q microresonators have become invaluable in applications requiring compact footprint, low cost, good energy efficiency, large comb spacing, and access to nonconventional spectral regions. In this review, we comprehensively examine the recent progress of photonic frequency microcombs and discuss how various phenomena can be utilized to enhance the microcomb performances that benefit a plethora of applications including optical atomic clockwork, optical frequency synthesizer, precision spectroscopy, astrospectrograph calibration, biomedical imaging, optical communications, coherent ranging, and quantum information science.

## 1. Introduction

Photonic frequency microcombs are sets of equidistant spectral lines generated via pumping a high-Q microresonator with a resonant continuous-wave single-mode laser. These microcombs have arguably created a new field in cavity nonlinear photonics, with a strong cross-fertilization between theoretical, experimental and technological research. It not only provides a new testbed for research in complex nonlinear dynamics, but also offers an alternative yet reliable route towards the long-sought-after vision of field-deployable optical frequency comb. Proof-of-principle demonstrations of microcomb based optical atomic clockwork, optical frequency synthesizer, precision spectroscopy, astrospectrograph calibration, biomedical imaging, optical communications, coherent ranging, and quantum information science have triggered great excitements in the science and engineering community. Here, we review the recent progress of photonic frequency microcombs based on dissipative Kerr soliton (DKS) and dissipative quadratic soliton (DQS) and discuss how various phenomena can be utilized to enhance the microcomb performances that benefit a plethora of applications.

The paper is organized as follows. Section [2] introduces DKS generation under linear and nonlinear effect, including modal coupling, spectral filtering, stimulated Raman scattering (SRS), nonlinear self-injection locking, stimulated Brillouin scattering (SBS), active gain inside and outside the microresonator, photorefractive effect, 2D material and quadratic nonlinearity. In Section [3], we focus on the theoretical work for pure DQS generation and corresponding dynamics, as well as providing experimental guidelines in both cavity-enhanced second-harmonic generation (SHG) and optical parametric oscillators (OPO). Soliton generation with coexisting quadratic and Kerr nonlinearities, is then discussed in Section [4]. Metrology tools to characterize the frequency combs, including the ultrafast soliton dynamics, soliton timing jitter and comb linewidth, are summarized in Section [5]. Finally, in Section [6], summary of rising and valuable applications of photonic frequency microcombs is given.

## 2. Photonic frequency microcombs based on dissipative Kerr cavity solitons

As an example of self-organization in driven dissipative nonlinear systems, dissipative Kerr soliton (DKS) frequency microcomb is generated by pumping a high-Q monolithic microresonator with a continuous-wave (CW) single-frequency laser. DKS formation critically relies on a double balance of nonlinearity and dispersion as well as dissipation and gain, and its dynamics is described by the Lugiato-Lefever equation (LLE) when high order effects are neglected [1,2]:

$$t_R \frac{\partial A}{\partial t} = -\left( \alpha + i\delta + i\frac{k^{"}L}{2}\frac{\partial^2}{\partial \tau^2} \right) A + i\gamma L |A|^2 A + \sqrt{\theta} A_p, \tag{1}$$

where $A$ is the intracavity field, $A_p$ is the amplitude of the incident pump field, $t_R$ is the round-trip time, $t$ is the slow time variable, and $\tau$ is the fast time that corresponds to the local time within the cavity, $\alpha$ is the total round-trip loss, $\delta$ is the detuning of the cavity resonance from the pump frequency, $L$ is the cavity length, $k^{"}$ is the group velocity dispersion (GVD) of the microresonator, $\gamma$ is the Kerr nonlinear parameter, $\theta$ is the pump coupling coefficient. The

LLE was first introduced in 1987 by Luigi Lugiato and Ren'e Lefever to analyze spatiotemporal pattern formation in a CW-driven, dissipative, diffractive and nonlinear optical cavity [1]. This mean-field equation combines the well-known nonlinear Schrödinger equation with the boundary conditions under the good cavity approximation [2]. Compared to coupled mode equations describing the mode interactions in frequency domain, LLE not only provides more direct insights into DKS dynamics in time domain but also provides a more straightforward way to include other nonlinear effects. LLE can be numerically solved by the split-step Fourier method, allowing computationally efficient modeling of octave-spanning microcombs even on a consumer-grade computer [2].

Early works on DKS microcomb focused on the nonlinear dynamics and spontaneous pattern formation governed by the LLE [Eq. (1)] as well as the experimental techniques to initiate and stabilize the broadband DKS microcombs [3]. After extensive study of the LLE-governed phenomena, research interests have been shifted to understanding and controlling more complex nonlinear cavity dynamics beyond the original LLE equation [Eq. (1)] as described in the remainder of this section. These studies not only deepen our knowledge of DKS microcomb, but also enable new ways to enhance the DKS microcomb performances that benefit a plethora of applications.

*2.1 Dissipative Kerr cavity soliton generation under linear and nonlinear effect*

*2.1.1 Linear effect*

A. Modal coupling

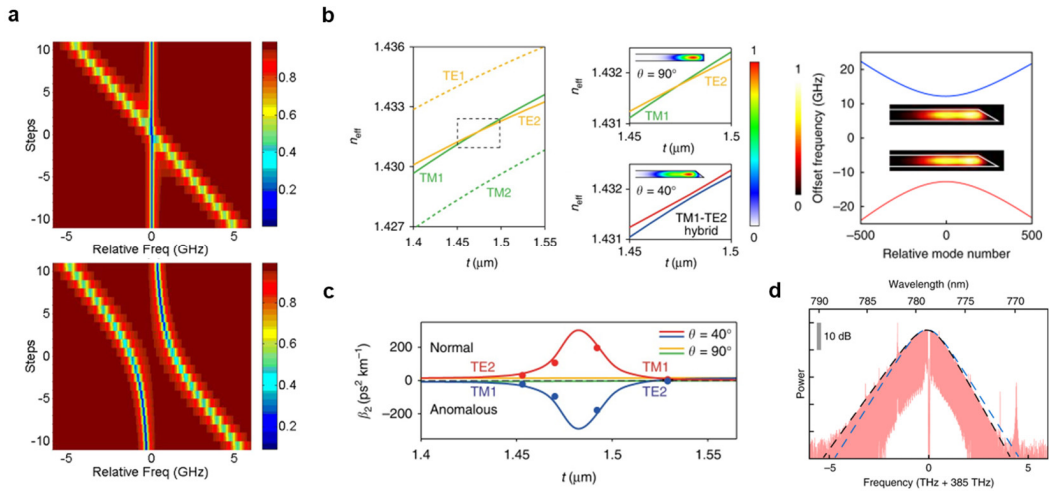

Fig. 1. Modal coupling engineering for DKS generation. (a) local dispersion due to weak and strong modal coupling [4]; (b) engineering of modal coupling through geometry control of microresonator [5,6]; (c) simulated and measured GVD with different geometries [5]; (d) Dirac soliton with asymmetrical optical spectra at visible band [6]. Reproduced with permission from Refs. [[4], [5], [6]].

In microresonators with multiple modes (either spatial modes or polarized modes), it is common that distinct mode families experience frequency degeneracy analogous to an energy level crossing, which terms as avoided mode crossings and are responsible for intriguing phenomena of both scientific and technical importance. In general, the modal crossings or modal couplings introduces local change of GVD (Figs. 1a and b) [4] or even reversed sign of GVD (Fig. 1c), leading to dispersive wave emission or DKS generation in nonconventional spectral regions (Fig. 1d).

Utilizing broadband modal coupling (Fig. 1b), dispersion engineering can be implemented to realize DKS especially in the visible band, where material dispersion or geometry-controlled dispersion alone is out of reach for anomalous GVD (Fig. 1c). DKS has been performed in a silica microresonator with equivalent anomalous dispersion at 778 nm and 1064 nm [5] in a broadband window, where the material GVDs are normal. In addition, DKS formed in this way is analogous to the Dirac soliton in the quantum field theory, which shows unusual properties, such as polarization twisting and asymmetrical optical spectra (Fig. 1d) [6].

Apart from broadband modal coupling effects, the more common narrowband modal coupling in anomalous-dispersion microresonators will result in spikes standing out of the soliton spectral envelope as well as dispersive wave emissions (also known as Cherenkov radiations) [7,8]. The emitted dispersive wave will lead to strong soliton perturbation and soliton interaction, including soliton crystal with high efficiency [9], breathing soliton [10], soliton annihilation and single soliton generation [11]. To satisfy the phase matching condition of Cherenkov radiation, soliton recoil is an evident effect with an overall shift of the soliton center frequency, leading to the inconsistency with the pump laser frequency. In addition, due to the soliton coupling to the dispersive wave, there exhibits hysteresis behavior in both the dispersive-wave power and the soliton properties, such as soliton center frequency shift and soliton repetition rate. In particular, the hysteresis behavior can reduce the soliton timing jitter at the quiet point wherein coupling of pump frequency noise into the soliton repetition rate is greatly weakened in various microresonators (Fig. 2a) [12,13]. However, there exists a limit of this noise suppression approach due to the frequency fluctuations of the dispersive wave, which is attributed to the intermodal thermal noise from unperfect mode overlapping between the dispersive-wave mode and soliton-forming mode (Fig. 2b) [14]. On the other hand, narrowband modal coupling in normal-dispersion microresonators leads to modulation instability (MI) in the confined spectral range where strong modal coupling locally change the sign of GVD for phase matched parametric oscillation. Such effect has been utilized to spontaneously excite dark DKS with high conversion efficiency and avoid Turing pattern destabilization for surpassing the conversion efficiency limit [4,[15], [16],[17]].

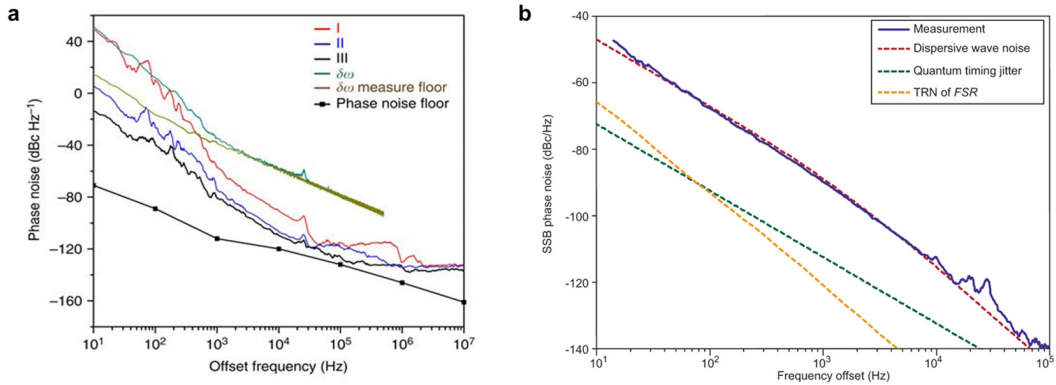

Fig. 2. Soliton timing jitter related to the dispersive wave. (a) soliton timing jitter reduction at different detunings (I, II and III) resulting from balance of dispersion wave induced soliton recoil and Raman induced self-frequency shift [12]. (b) noise limit imposed by inter-modal dispersive wave [14]. Reproduced with permission from Refs. [12,14].

In addition to the above-mentioned mode coupling between different mode families in a single microresonator, mode interactions can also be realized in multiple coupled microresonators, where the coupling strength can be easily adjusted [18,19]. By controlling mode interactions of dual-coupled microresonators with on-chip integrated heaters, repetition-rate selectable and mode-locked dark-soliton combs are realized in normal-dispersion microresonators [20]. The microcomb is generated only in the pumped main microresonator due to the large free spectral range (FSR) difference induced asynchronization between the coupled microresonators. The coupling strength is found to be important for expanding the existence range of dark-soliton in the pump-power-detuning bidimensional space. Dispersion engineering and dispersive wave emissions are also expected in this type of coupled photonic molecules.

B. Spectral filtering

In microresonators with anomalous GVD, both experimental and numerical results show that a strong and localized spectral loss may completely stop the generation of DKS combs [21], depending on the loss bandwidth and location. Similar to above mentioned dispersion anomalies resulting from the modal couplings, the effect of bandpass intensity modulation can introduce MI and frequency combs even with normal dispersion by the mechanism of gain-through-filtering [22]. Moreover, one of the authors in *our group* finds that the bandpass filtering is responsible for generating and stabilizing chirped ultrashort pulse in a normal-dispersion $Si_3N_4$ microresonator, where the spectral filtering is achieved by the intrinsic overtone absorption from the surface N–H bonds (Fig. 3a) [23]. Besides, chirped pulse generation is also experimentally demonstrated with an optical bandpass filter in all-normal-dispersion fiber based passive resonator [24]. In contrast to the conventional DKS in anomalous-dispersion and high-Q microresonators, the experimental and numerical results [25] indicate various mode-locked states which are insensitive to the system dissipation and can be attributed to the strong intracavity pulse evolution due to the normal dispersion and spectral filtering. The chirped pulse generation with spectral filtering in passive resonators (Fig. 3b) provides an evident complement to its counterpart of the all-normal-dispersion mode-locked fiber laser [26].

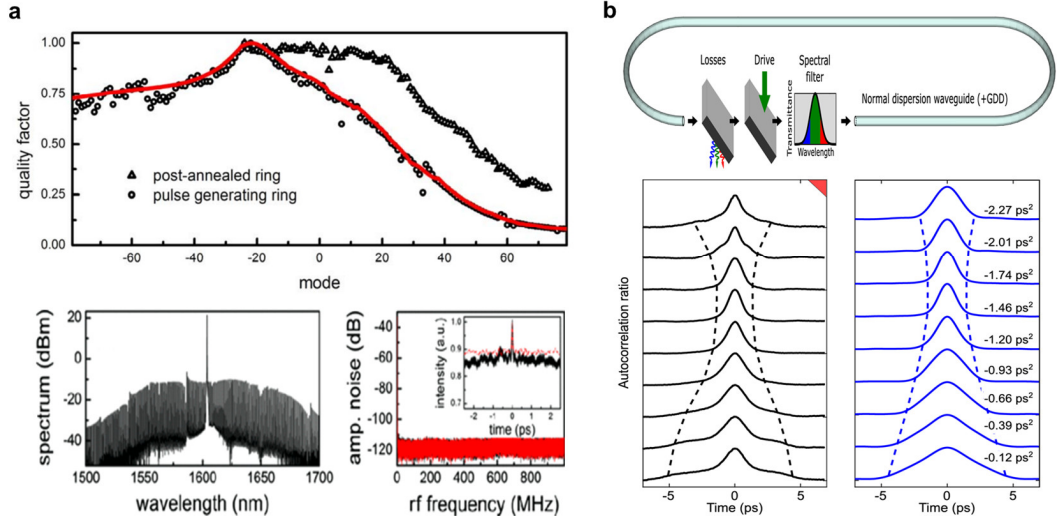

Fig. 3. Spectral filtering stabilizes mode-locking in microresonators with normal dispersion. (a) weak normal dispersion and weak spectral filtering effect [23]; (b) chirped pulse generation with strong spectral filtering effect [24]. Reproduced with permission from Refs. [23,24].

*2.1.2 Nonlinear effect*

A. Stimulated Raman scattering (SRS)

In any molecular medium, Raman scattering can transfer energy from one to another optical field, whose frequency is downshifted by an amount determined by the vibrational modes of the medium. In addition, Raman scattering can occur within an ultrafast pulse itself, where the blue spectral components serve as the pump and amplify the red spectral components. For DKS in microresonators, intrapulse Raman scattering can lead to soliton self-frequency shift, depending on the dispersion, laser detuning and Raman shock time [27]. Besides, weak Raman effect is theoretically predicted to introduce MI in microresonators with normal dispersion and facilitate dark soliton generation [[28], [29], [30]].

However, strong Raman scattering usually prevents DKS generation due to the competition between Raman and Kerr effect [31] and large portion of energy transferred from pump to the Raman modes. In order to alleviate Raman effect and achieve DKS, several methods can be implemented: (i) reducing the microresonator size with large FSR to stagger the Raman gain peak and resonant peaks [32]; (ii) orientating field polarization along the proper crystal axis to reduce the Raman gain; (iii) increasing the loss of Raman mode by dissipative control, through for example auxiliary microrings, engineered pulley waveguide couplers, scattering centers and self-interference coupling structure (Fig. 4) [33].

Despite the perturbation, strong Raman effect can also be utilized to generate mode-locked Raman soliton [34] in microresonators, similar to the counterpart in fiber lasers [[35], [36], [37]]. Three conditions must be fulfilled for bright Raman soliton in microresonators: (i) anomalous GVD for the Raman mode family; (ii) broadband Raman gain created by the pump soliton; (iii) small group velocity difference between the pump soliton and Raman soliton. With these conditions fulfilled, the Raman soliton will trap and propagate together with the pump soliton at a same group velocity. Usually, the third condition can not be satisfied in the same spatial mode due to the large group velocity modification across a large Raman frequency shift. Therefore, two different spatial modes for pump soliton and Raman soliton respectively [34] are introduced to achieve small group velocity difference and soliton trapping via self-phase modulation and cross-phase modulation.The Raman soliton can be viewed as a transferred soliton from the pump soliton, experiencing both Raman gain and parametric gain. While careful dispersion engineering is required in few-mode microresonators to fulfill the three conditions, it is of advantage to achieve Raman soliton in over-moded microresonators such as microspheres [38]. In all, the Raman soliton provides an efficient method for background-free soliton with extended longer wavelength.

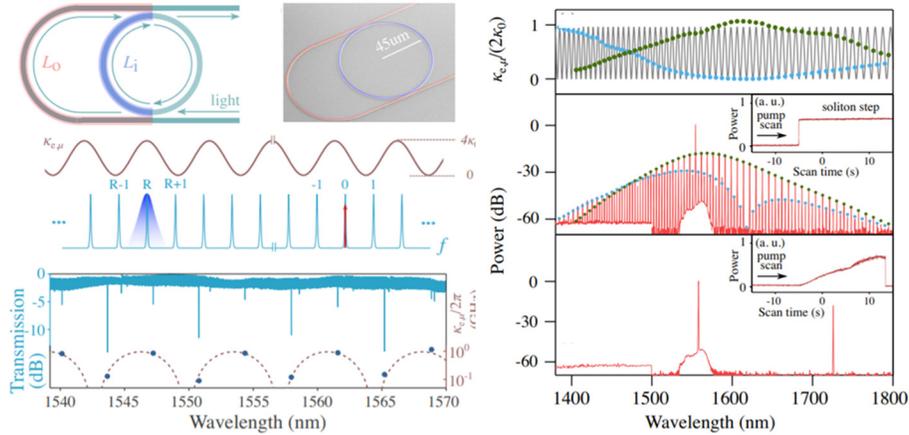

Fig. 4. Photonic dissipation control to achieve DKS by inducing extra loss for Raman modes within the same mode family [33]. Reproduced with permission from Ref. [33].

B. Nonlinear self-injection locking

Self-injection locking (SIL) with a high-Q microresonator as the feedback element was first proposed to significantly reduce the linewidth of a single-frequency laser, either a semiconductor laser [[39], [40], [41]] or a fiber laser [42]. When the laser frequency coincides with the resonant frequency of the microresonator, the laser couples into the microresonator. If some amount of the light in the microresonator is fed back into the laser cavity through for example Rayleigh scattering [43], the laser frequency can be locked to the microresonator resonance with real-time compensation of the thermal effects. Benefitting from the fast response, by fine tuning the laser frequency within the locking bandwidth, DKS can be stably accessed in the red-detuned regime without compensating the thermal nonlinearity using active electronics. Of note, the precise control of laser frequency can be achieved via tuning drive current and the phase depending on both feedback path and the nonlinear phase from self- and cross-phase modulation. The microresonator plays two roles: one is the feedback element to narrow the laser linewidth and the other is a nonlinear medium for DKS generation. Therefore, nonlinear self-injection locking method for DKS generation eliminates the requirement of the active electronics for DKS initiation and stabilization, compared with conventional DKS generation, rendering it a promising method to integrate microresonators and chip-scale lasers on an ultracompact single chip.

By implementing the nonlinear self-injection locking principle, generation of stable Kerr frequency combs based on dark soliton and bright soliton were first realized in $MgF_2$ crystal microresonators [44,45]. These experiments show the ability to simultaneously achieve both ultralow soliton timing jitter and ultranarrow comb linewidth inherited from the locked laser. Usually, a single-frequency DFB laser with relatively broad linewidth at free running status is introduced for SIL. However, ordinary cost-effective Fabry-Pérot (FP) laser diodes have also been proved to be feasible [46], with the additional advantage of higher output power compared with DFB lasers.

Discrete laser chips and high-Q on-chip microresonators are firstly demonstrated to show the potential for the integration of DKS system [43,47,48]. Ultrahigh-Q microresonators are required to lower the soliton threshold since the laser chip output power is usually limited at ~40 mW at telecom wavelengths. To integrate the DKS system on a COMS-compatible platform, photonic Damascene reflow process [43] is introduced to boost the Q factor of $Si_3N_4$ microresonators and lower the Kerr parametric oscillation threshold to ~1 mW level. By directly butt-coupling a multi-frequency Fabry–Pérot laser diode chip or a DFB laser chip onto a $Si_3N_4$ photonic chip, integrated turnkey DKS microcomb system is realized in a compact butterfly package [43]. For further stability and minimize of the coupling loss, both indium phospide/silicon (InP/Si) semiconductor lasers and $Si_3N_4$ microresonators are heterogeneously integrated on a monolithic silicon substrate (Fig. 5) [49]. In addition, both the soliton timing jitter and the comb linewidth are efficiently reduced due to the self-injection locking [48]. These landmark works [43,[47], [48], [49], [50], [51], [52]] open the route towards high-volume production and practical applications.

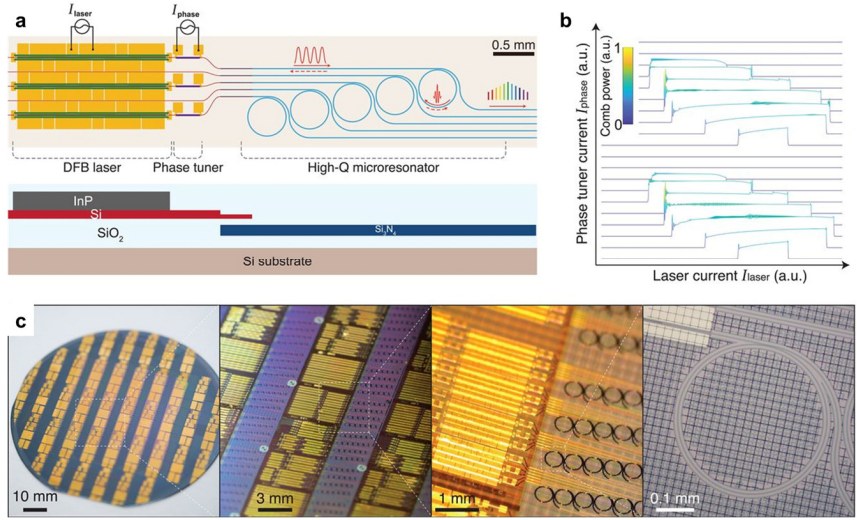

Fig. 5. Integration of both semiconductor lasers and microresonators on a monolithic silicon substrate [49]. Reproduced with permission from Ref. [49].

C. Stimulated Brillouin scattering (SBS)

SBS lasers have been demonstrated in high-Q microresonators with ultranarrow linewidth and ultralow threshold [53,54], benefiting applications that require ultrahigh laser coherence. Recently, *our group* showed that controlled interaction between SBS and Kerr nonlinearity results in generation of self-stabilized microcombs with ultranarrow comb linewidth and ultralow DKS timing jitter [[55], [56], [57], [58]]. The demonstrations were first done in a monolithic highly nonlinear fiber Fabry-Perot (HNLF FP) resonator [55], and later extended to graded index multimode fiber Fabry-Perot (GRIN-MMF FP) resonators [[56], [57], [58]]. In these experiments, we introduced a novel two-step pumping scheme where the primary pump at one mode family triggers SBS lasing that serves as the secondary pump for DKS generation at another mode family. The blue-detuned primary pump and the red-detuned SBS secondary pump can work together to compensate the detrimental cavity thermal nonlinearity that limits the reliability and robustness of DKS generation, with the same principle used in the auxiliary-assisted microcavity method [59]. The two-step pumping scheme also takes advantage of SBS laser's orders-of-magnitude linewidth reduction [53,54] to not only narrow the microcomb linewidth but also lower the DKS timing jitter by mitigating the detuning noise [55]. In the first demonstration with HNLF FP resonator, the microcomb linewidth of 22 Hz and quantum-limited DKS timing jitter of 995 as for averaging times up to 10 μs were achieved (Fig. 6a) [55]. Such ultralow timing jitter can be measured by an all-fiber reference-free Michelson interferometer (ARMI) timing jitter measurement apparatus as described in Section 5.3 [60]. Of note, the two-step pumping scheme has also been applied to other DKS platforms such as silica microdisk resonator [61] and silica wedge resonators [62] by other groups. It should also be applicable to Raman-Kerr soliton generation in crystalline microresonators with narrowband Raman gain such as aluminium nitride (AlN) microresonators [63].

By implementing the two-step pumping scheme between two spatial modes in a GRIN-MMF FP resonator, *our group* recently demonstrated unprecedented microcomb linewidth of 400 mHz and DKS timing jitter of 500 as for averaging times up to 25 μs (Fig. 6b) [57]. The enhanced performances are attributed to the improved Q factor, lower thermos-refractive noise [64], and lower quantum limit [65] of the GRIN-MMF FP resonator. In addition, GRIN-MMF's low modal dispersion enhances the intermodal nonlinear interaction [66] and by leveraging such property *our group* was able to recently demonstrate the generation of spatiotemporal mode-locked DKS for the first time [57].

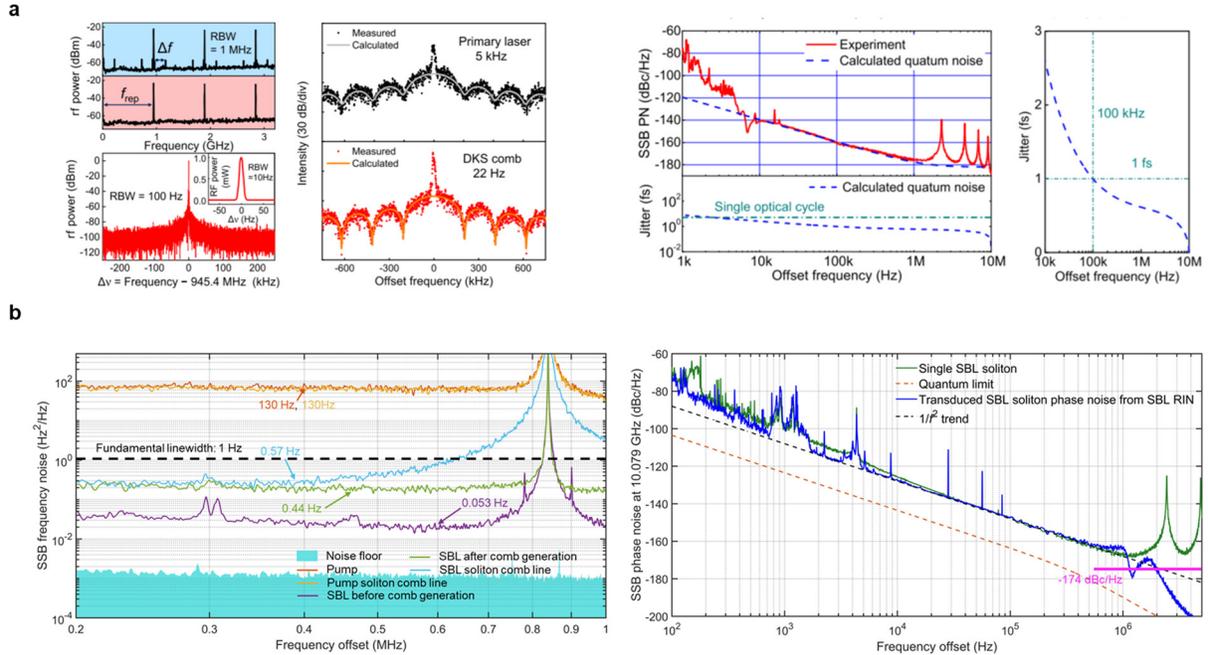

Fig. 6. Characteristics of Brillouin-Kerr comb including microcomb linewidth and DKS timing jitter. (a) SBS between two modes with orthogonal polarizations [55]; (b) SBS between two spatial modes [57]. Reproduced with permission from Ref. [55].

D. Active gain effect inside the microresonator

Active gain material can be incorporated to increase the Q factor and lower the pump threshold for DKS generation. Of note, these microresonators are configured to have gains close to the lasing threshold and no saturable absorber, fundamentally different from mode-locked lasers. Such configuration ensures that DKS generation is still in the conventional framework where DKS is coherently driven by the external CW pump. The first demonstration is done in a long class B erbium-doped fiber cavity [67], showing that spontaneous soliton formation is forbidden with a CW external pump but becomes accessible with a pulsed external pump. The most significant finding is that, to the limit of the equipment, the so-called active cavity soliton does not suffer from the Gordon-Haus jitter resulting from the amplified spontaneous emission. Later, it is shown that a generalized LLE can also be used to describe the dynamics of a class B quantum cascade laser (QCL) with an external pump under weak saturation [68]. Thus, localized structures such as Turing pattern and DKS in externally pumped QCL was predicted and numerically analyzed [68,69]. Surprisingly, the first DKS generation in QCL was demonstrated without any external pump [70], and gain saturation may be a reason why the results are different from previous theoretical analyses [71]. Without an external pump, higher efficiency can be achieved, and it is more straightforward to extend the DKS wavelength range to mid-infrared and terahertz. Despite the remarkable results, more studies are necessary to better understand its physical mechanism, how it differs from saturable absorber enabled mode-locking, and what the consequences are.

E. Laser with a nested microresonator

Fig. 7b depicts an erbium-doped fiber cavity with a nested microresonator where the microresonator provides not only narrow linewidth filtering but also enhanced Kerr nonlinearity [[72], [73], [74]]. This filter-driven four-wave mixing laser [73] is fundamentally different from the nonlinear self-injection locking (Section B) in that the microresonator is now an integral part of the laser rather than an element that provides feedback to the laser. Besides high Q factor, the nested microresonator should be designed to also have a high transmission to lower the lasing threshold and thus add-drop type microresonator is usually preferred in this configuration. A delay line is introduced to the faber cavity to fine tune the relative position of the lasing modes with respect to the microresonator resonances, which is critical to achieve the generation of the so-called laser cavity soliton (LCS) [74]. Self-starting background-free solitons with a mode efficiency of 75%, a few times higher than conventional DKS, was demonstrated [74].

In our recent study [75], we theoretically showed that a red-detuned nested microresonator has a nonlinear transmission that provides an effective saturable absorber action to mode-lock the fiber laser. The insight leads to the prediction of a new class of bright chirped LCS and the demonstration of a mode efficiency as high as 90.7%, approaching the theoretical limit of 96% [75]. In addition, intriguing LCS interaction phenomena including optical Newton's cradle were experimentally observed [75]. Despite the remarkable results, more studies are necessary to better understand the physical mechanism that can expand the LCS bandwidth, lower its phase noise, and improve its long-term stability for practical applications.

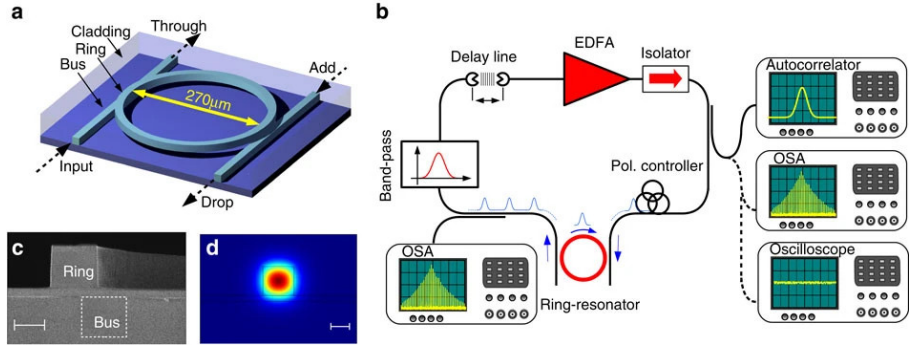

Fig. 7. Demonstration of a stable ultrafast laser based on a nonlinear microcavity. Reproduced with permission from Ref. [73].

F. Photorefractive (PR) effect

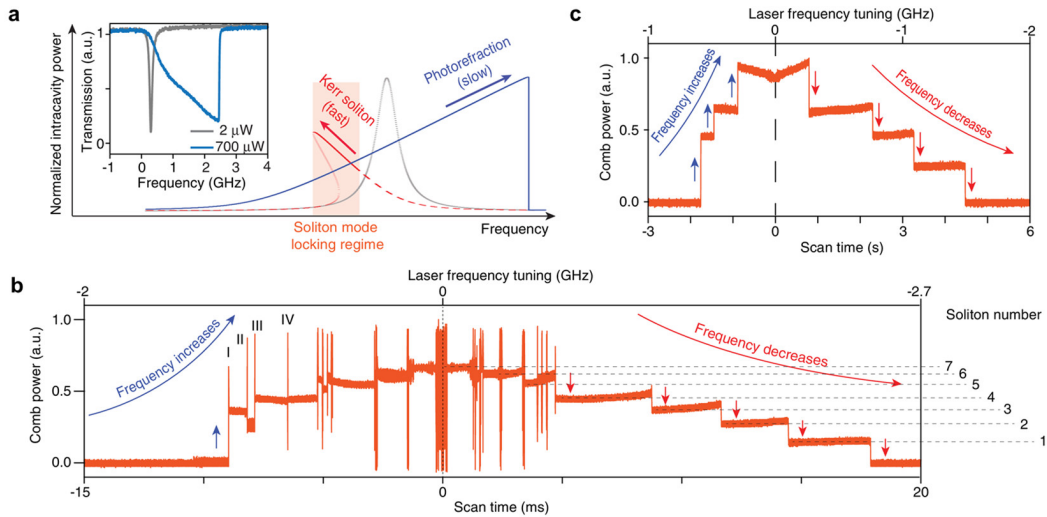

Fig. 8. PR effect on DKS dynamics. (a) influence of PR effect on cavity transmission; (b) bi-directional soliton switching with a fast pump frequency scanning speed; (c) bi-directional soliton switching with a slow pump frequency scanning speed. Reproduced with permission from Ref. [76].

For most of the microresonators, DKS exists in the red-detuned regime where exhibits thermal instability due to the thermal-optic effect and causes an average-power-dependent increase in the refractive index. The detrimental thermal nonlinearity prohibits the soliton access and requires complex pump tuning techniques to trigger soliton generation. In contrast to the thermal-optic effect, the PR effect causes an intensity-dependent decrease in the refractive index, which can be utilized to compensate the detrimental thermal-optic effect, make the red-detuned regime thermally stable and self-start the DKS. As an example of microresonator on a Z-cut $LiNbO_3$ thin film [76], the PR effect on the refractive index is dominant over the weak thermo-optic effect. The response time of PR effect is much slower than that of the thermo-optic effect and the Kerr effect (Fig. 8a). Bi-directional soliton switching is observed deterministically by tuning the pump wavelength at a speed matching the response time of PR effect. With pump frequency up tuning from the red side, the solitons first self-start and then switching to high-number state by re-entering the MI regime (Fig. 8b). The final soliton number results from the balance between the three effects. The mediated MI process can be indicated by spikes on the average power traces (Fig. 8b). With slower tuning speed, spikes will disappear (Fig. 8c). The soliton under pump frequency down tuning behaves in a similar manner to the conventional DKS. In addition, the PR effect helps to achieve perfect soliton crystal on demand [77]. The stable access for DKS and convenient DKS switching resulting from PR effect, can contribute to device integration for practical applications by removing electronics for soliton initiation and stabilization.

G. 2D material

2D materials are widely applied in lasers especially for novel mode-locked lasers, due to their unique optoelectrical responses, for example the fast saturable absorption (SA). Triggered by the same idea, deterministic and self-starting localized structures including DKS are theoretically predicted to exist at close-to-zero detuning values with SA effect under external drive [78,79]. However, few experimental results are demonstrated by combining 2D materials and high-Q microresonators. Reasons can be attributed to (i) the difficulty of depositing 2D materials onto the monolithic microresonators; (ii) the low damage threshold of 2D materials; (iii) the Q-factor reduction due to the deposition.

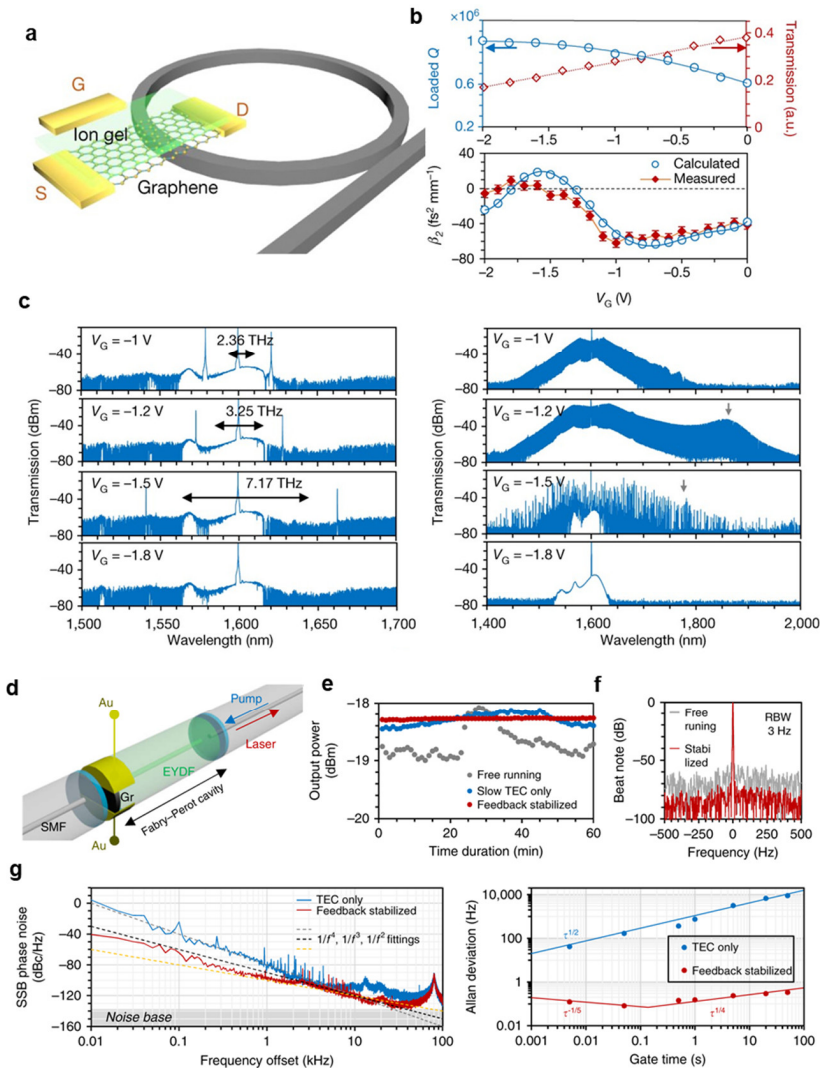

Fig. 9. Microcombs combined with 2D materials. (a)(b)(c) Gate-tunable frequency combs in graphene–nitride microresonators [80]. (a) Schematic architecture; (b) Gate-tunable dispersion and corresponding Q factors; (c) Gate-tunable frequency comb. (d)(e)(f)(g) Electrically controllable laser frequency combs in graphene-fiber microresonators [81]. (d) Schematic architecture. Stabilized power (e), repetition rate (f)(g) and Allan deviation (g). Reproduced with permission from Refs. [80,81].

In spite of these difficulties, remarkable results have been demonstrated in microresonators with electrically controlled graphene. The first demonstration [80] is from one of the authors in *our group* where the second- and higher-order chromatic dispersions of a CW-driven silicon nitride microresonator are actively modulated by coupling gate-tunable optical conductivity and altering Fermi-level of graphene (Fig. 9a). The interacting graphene is top-gated for reducing photon absorption in the nearly massless Dirac cone and keeping the high-Q factor of the microresonator. As shown in Fig. 9b, the GVD of the microresonator can be tuned in a large range while keeping the high-Q factor. The tunable dispersions through gate voltage, result in the tuning of generated combs, including primary comb line locations, the full comb bandwidth and the frequency spacing between the pump and the Cherenkov radiation (Fig. 9c). This heterogeneous graphene microcavity, which combines single-atomic-layer nanoscience and ultrafast optoelectronics, not only improves the understanding of dynamical frequency combs and ultrafast optics, but also the benefits the applications of microcomb.

Another example [81] is the electrically controllable laser frequency combs in a heterogeneous graphene-fiber microcavity (Fig. 9d). Despite the same idea of mode-locked fiber laser where graphene serves as the SA, the electrically controlled Fermi level of atomically thick graphene, leads to tunable absorption, controllable Q-factor, and fast optoelectronic feedback stabilization. As for the comb characteristics, tunable repetition rates, controllable wavelengths, and self-stabilized phase noise (Fig. 9e–g) are demonstrated for ultracompact fiber laser combs with high repetition rates.

H. Quadratic nonlinearity

The coexisting Kerr nonlinearity and quadratic nonlinearity in the same microresonator can trigger many intriguing phenomena and corresponding applications, including parametrically driven DKS, dark soliton, deterministic single DKS and monostable DKS with self-starting behavior. To better organize the review, details can be found in Section 4.

### 2.2 Multistability and multiple combs generation in a single microresonator

In Eq. (1), DKS is spontaneously formed from the four-wave-mixing induced MI in the bistable regime. Recent experimental demonstrations imply multiple solitons formation in the multi-stable regime, which triggers multiple comb generation and corresponding applications. Multistability in a single microresonator, can result from interplaying between modes with orthogonal polarizations, different spatial modes, counter-propagating modes, different wavelengths and giant nonlinearity across adjacent resonances. Different from systems with single-mode comb generation, group velocity mismatch (GVM) and cross-phase modulation (XPM) effects must be considered in systems with multi-stability.

Ref. [82] reports the orthogonally polarized frequency comb generation from DKS comb in a $Si_3N_4$ microresonator. The DKS comb results from the TE mode with anomalous dispersion, while the other combs at the TM mode are generated via XPM. No soliton is formed for the TM mode due to its normal dispersion. However, things go different in the case of orthogonally polarized mode with both anomalous dispersions. Ref. [83] experimentally demonstrates coexisting solitons with different spectral widths for orthogonally polarized modes in a fiber resonator (Fig. 10a), which is numerically predicted in Ref. [84]. In addition, multi-stability is also observed, such as coexistence between the following states: (i) a near-periodic and stable MI pattern and a DKS, (ii) an aperiodic and unstable MI pattern and a DKS. To achieve the coexistence of these nonlinear states, the two resonances should be adjusted by birefringence to be close to each other such that they can be simultaneously pumped. Moreover, GVM between the two polarized modes should be kept small to mutually trap the two solitons. DKS combs with different repetition rate is not realized in Ref. [84] perhaps due to the strong XPM effect and the small GVM.

By introducing multiple narrow-linewidth pump lasers or frequency shifters, it is flexible to study the multi-stability associating with various resonant spacings. Ref. [85] demonstrates the first spatial multiplexing soliton microcombs, where solitons coexist between two or three spatial modes. Thanks to the reduced XPM effect and alleviated inter-locking effect resulting from small spatial overlapping, dual DKS combs are demonstrated with repetition rate differences of 655 kHz and 9.3 MHz with FSR of 12.4 GHz. Generated from two pumps via a frequency shifter, the dual-comb is inherently coherent and readily for applications such as spectroscopy and resolving the breathing dynamics of a soliton. In addition, the authors also demonstrated the ability to multiplex three soliton combs by pumping three mode families simultaneously, in which two mode families are co-pumped in the clockwise direction, while another comb is generated by pumping a third mode family in the counter-clockwise direction.

Dual-comb generation can also be realized by counter-propagating (CP) solitons with the same spatial mode in a single microcavity [86]. Of note, resulting from the intra-cavity back-scattering and XPM effect, the relative motion of clockwise and counter-clockwise solitons can be adjusted and switched between locked and unlocked states by changing the counter-pump frequency difference. The almost linear relationship between the repetition rate difference and the counter-pump frequency difference makes it a flexible and powerful tool for dual-comb spectroscopy (see Section 6.5). Besides, by utilizing a balanced optical cross-correlator (BOC) (see Section 5.3), the relative timing jitter of the CP solitons is measured to reach the quantum limit [87].

It is also possible to achieve multi-stability enabled composite solitons by pumping different resonances within the same spatial mode in a co-propagating way. For example, Weng et al. observes composite solitons exhibiting complex frequency comb patterns and successive soliton collisions. These states are achieved by pumping a crystalline microresonator with two lasers that are frequency detuned from each other by one or multiple cavity FSRs [88]. Reversible transition from repetition-rate synchronized solitons to unsynchronized solitons and dual-comb is clearly demonstrated via electro-optic sampling technique (see Section 5.1) [89]. With the same method, the authors also explore the multi-stability regime by pumping the same resonance with two frequency-detuned lasers. In this case, heteronuclear soliton molecules consisting of composite solitons are achieved (Fig. 10b) [90]. Differently, these solitons share the same repetition rate and lock to each other.

The bichromatic pumping scheme can be further extended for broadband comb spectrum. Ref. [91] reports the spectral extension from 1300 nm to 1700 nm via primary and auxiliary pump lasers. The primary soliton comb is generated around 1.5 μm for a mode family with anomalous dispersion, while the auxiliary comb is generated around 1.3 μm for the same or different mode family with normal dispersion (Fig. 10c). Synchronization between these two combs is realized by XPM effect. Besides, the bichromatic pumping scheme is proved to effectively alter the dispersion landscape through non-degenerate four-wave mixing enabled dispersive wave generation, which creates Kerr soliton microcombs far beyond the intrinsic anomalous dispersion regime [93]. This concept of synthetic dispersion is applied to generate a single-soliton microcomb whose bandwidth approaches two octaves (137 THz to 407 THz) by pumping a $Si_3N_4$ microring resonator at 1060 nm and 1550 nm. Of note, the bichromatic pumping scheme is different from the soliton stabilization method using an auxiliary laser where the auxiliary laser does not participate in the comb generation.

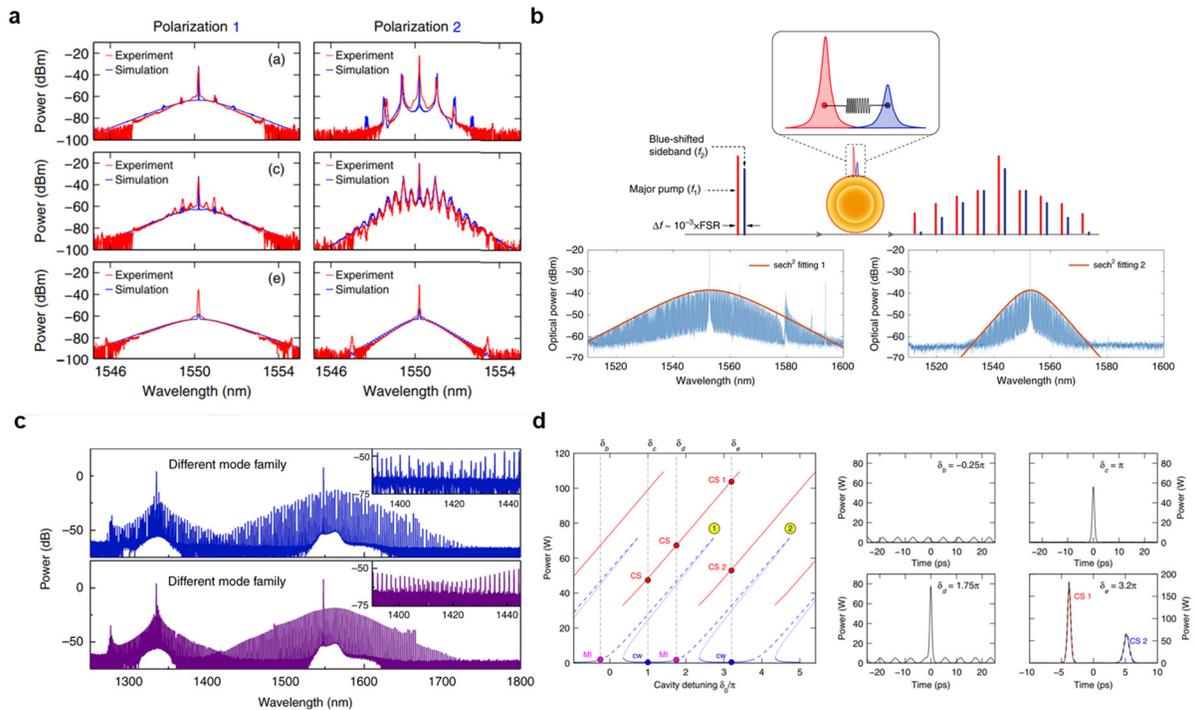

Fig. 10. Multiple solitons generation in a single microresonator. (a) coexisting soliton with orthogonal polarizations [83]; (b) heteronuclear soliton molecules consisting of composite solitons [90]; (c) spectral extension and synchronization of microcombs through bi-chromatic pump [91]; (d) Super cavity soliton generation [92]. Reproduced with permission from Refs. [83,[90], [91], [92]].

Apart from the above-mentioned bi-pumping schemes, multistability can be realized with single pump laser by overlapping two nonlinearly tilted and adjacent cavity resonances (Fig. 10d). This method requires giant Kerr nonlinearity and relatively small cavity FSR. In Ref. [92], the authors synchronously drive two single-mode fiber ring resonators (FSR of 2 MHz and 0.7 MHz) to experimentally access this regime and observe the rise of new nonlinear dissipative states, including the extended MI (Turing) patterns and the coexistence of two distinct cavity soliton states, one of which can be identified as a "super" cavity soliton [[94], [95], [96]].

In addition, multistability can also occur in the low quality-factor fiber cavities by synchronous pumping, producing chirped solitons [24].

*2.3 Multiple combs generation in multiple coupled microresonators*

Despite the multiple solitons generation in a single microresonator, it is important to explore soliton generation in multiple coupled microresonators which can offer more freedoms, such as variable coupling strengths and flexible cavity lengths and dispersion engineering. The additional microresonators provide a beneficial testbed to study the soliton interacting behaviors and trigger interesting applications.

The first example is the synchronization of DKS in $Si_3N_4$ microresonators on separated chips [97]. DKSs are first generated with the same pump laser in two near identical microresonators. A small portion of the master soliton power (1%) is injected into the slave soliton through a 20-m fiber. By fine adjusting the fiber length and the microresonator size through on-chip heater, synchronization of two DKSs is achieved, which is advantageous for the enhancement of the nonlinear conversion. At the synchronized state, spectral fringes in the spectrum of the slave comb are observed due to the stable interference between the slave comb and sync signal. In addition, synchronization of weakly coupled optical microresonators with different sizes is also achieved, resulting in Arnold tongues [98].

With strong linear coupling between two identical microresonators, emergent nonlinear dynamics is experimentally observed and theoretically analyzed [99]. In this case, single pump is coupled into one microresonator while DKSs can be generated in both microresonators. Novel and interesting phenomena are observed, ranging from gear soliton generation, symmetry breaking, resonant Fano dispersive waves and soliton hopping. Note that the soliton generation in both coupled microresonators is fundamentally different from frequency comb generation with another size-mismatched auxiliary microresonator served as pump booster [100] or MI initiator [20].

In addition to the two coupled microresonators, pattern formation and nonlinear dynamics are theoretically investigated in an array of equally coupled Kerr microresonators [101], which are associated with an effective two-dimensional space. Global nonlinear optical patterns are demonstrated, which correspond to coherent optical frequency combs on the individual resonator level.

Besides, complex arrays of hundreds of coupled ring resonators usually lead to topological photonic systems. Remarkably simple features can exist in the topological systems, such as edge states, which are dictated only by the global topology and therefore are independent of local details of the system. This unique property of edge states protects them against local defects and disorders in the system, enabling the realization of robust photonic devices. Ref. [102] theoretically investigates the generation of coherent optical frequency combs and temporal DKSs in a topological photonic system consisting of a two-dimensional lattice of coupled microring resonators. Topological edge states are found to circulate around the boundary of the lattice and a travelling-wave super-ring resonator is formed from multiple single-ring resonators. By pumping the super-ring resonator with a CW laser, temporal patterns such as Turing rolls and nested DKSs, are spontaneously formed. These temporal patterns are phase-locked across all the ring resonators on the edge of the lattice. In particular, the nested solitons inherit the topological protection of a linear system and are robust against any defects in the lattice. Another advantage is the high mode efficiency of >50% for a single nested soliton, which is an order of magnitude higher than the theoretical efficiency of ∼5% for conventional DKS combs.

**3. Photonic frequency microcombs based on dissipative quadratic cavity solitons**

Recent experiments [[103], [104], [105]] have demonstrated that frequency combs can also be entirely generated through the second-order (quadratic) nonlinearity. Modulation instability is found in either cavity-enhanced SHG [103] or degenerate optical parametric oscillator (OPO) [104] using periodically poled lithium niobate (PPLN) crystal operating at 1064 nm and 532 nm. The frequency comb generation is attributed to the nonlocalized nonlinearity depending on the system parameters, including the quadratic nonlinearity, resonance detuning, the phase mismatch and the dispersions. There are growing interests to go beyond MI induced quadratic frequency comb towards DQS generation for higher coherence, broader bandwidth, and better pulse quality. Due to the orders-of-magnitude higher quadratic nonlinearity than the Kerr nonlinearity, DQS tends to have a lower pump threshold and higher pump-to-soliton conversion efficiency compared with DKS [106]. In addition, DQS provides a feasible solution to extend the microcomb wavelength to difficult-to-access spectral ranges including the midinfrared molecular fingerprinting region and ultraviolet region. Finally, the quadratic nonlinearity in DOPOs can add another degree of freedom to the soliton solutions and enable intriguing applications including random bit generation [107] and optical Ising machine [108].

Despite the promising outlook, there are few experimental realizations of quadratic frequency comb, let alone DQS. All the experimental results achieved so far are based on the cooperation between $\chi^{(2)}$ and $\chi^{(3)}$ nonlinearities, which will be discussed in Section 4. Here, we will review and summarize the recent theory development on DQS based OFCs [[109], [110], [111], [112], [113], [114], [115], [116], [117], [118], [119], [120], [121], [122], [123]], and more importantly provide a general picture of the DQS existence and characteristics as well as the experimental guidelines for DQS generation based on *our group*'s extensive and systematic DQS study in the recent years [[121], [122], [123]]. We first simplify the system of equations that describes the quadratic nonlinear cavity dynamics into a single mean-field equation so that the effects of cascaded quadratic process on the DQS become more transparent for gaining insights and performing analyses. Next, we further simplify the single mean-field equation into a form similar to the well-known LLE or parametrically driven nonlinear Schrödinger equation (PD-NLSE) to not only greatly reduce the computational complexity for solving the equations, but also provide more insights about the DQS existence, dynamics, and property.

*3.1 Mean-field equation for quadratic soliton*

*3.1.1 Single mean-field equation*

As shown in Table 1, four cases can be classified according to the different boundary conditions, including singly resonant SHG cavity (SR-SHG), doubly resonant SHG cavity (DR-SHG), singly resonant degenerate OPO (SR-DOPO) and doubly resonant degenerate OPO (DR-DOPO). Two-wave coupled equations together with the boundary conditions [103,104], are required to precisely describe the nonlinear cavity dynamics dominated by the second-order nonlinearity. Fortunately, the system of equations can be simplified into a single mean-field equation under the high-Q cavity assumption which is valid for microresonators studied here (Table 1). Singly resonant cases correspond to nonlinear response function $I(\tau)$ [116,122], while doubly resonant cases correspond to $J(\tau)$ [121], respectively. For all the mean-field equations, the fourth term on the right-hand side exhibits an effective third-order nonlinearity with a non-instantaneous response, which is similar to the delayed Raman response found in cubic Kerr media and other non-instantaneous nonlinear Schrodinger models. However, contrary to those models there is also a phase dependence resulting from the square of the field rather than the intensity. Of note, to draw readers' attention on the main conclusion, the related symbol definition can be found in Appendix.

**Table 1. Four cases determined by the boundary conditions and corresponding single mean field equation.**

| Case | Description | Single mean-field equation |
|---|---|---|
| SR-SHG | driven at $\omega_0$, resonant at $\omega_0$ | $t_R \dfrac{\partial A(t,\tau)}{\partial t} = -\left(\alpha_1 + i\delta_1 + i\dfrac{k_1^{''}L}{2}\dfrac{\partial^2}{\partial \tau^2}\right)A - (\kappa L)^2 A^*\left[A^2(t,\tau) \otimes I(\tau)\right] + \sqrt{\theta_1}A_{in}$ |
| DR-SHG | driven at $\omega_0$, resonant at $2\omega_0$ and $\omega_0$ | $t_R \dfrac{\partial A(t,\tau)}{\partial t} = -\left(\alpha_1 + i\delta_1 + i\dfrac{k_1^{''}L}{2}\dfrac{\partial^2}{\partial \tau^2}\right)A - \left[\kappa L\operatorname{sinc}(\xi/2)\right]^2 A^*\left[A^2(t,\tau) \otimes J(\tau)\right] + \sqrt{\theta_1}A_{in}$ |
| SR-DOPO | driven at $2\omega_0$, resonant at $\omega_0$ | $t_R \dfrac{\partial A(t,\tau)}{\partial t} = -\left(\alpha_1 + i\delta_1 + i\dfrac{k_1^{''}L}{2}\dfrac{\partial^2}{\partial \tau^2}\right)A - (\kappa L)^2 A^*\left[A^2(t,\tau) \otimes I(\tau)\right] + i\mu A^*$ |
| DR-DOPO | driven at $2\omega_0$, resonant at $2\omega_0$ and $\omega_0$ | $t_R \dfrac{\partial A(t,\tau)}{\partial t} = -\left(\alpha_1 + i\delta_1 + i\dfrac{k_1^{''}L}{2}\dfrac{\partial^2}{\partial \tau^2}\right)A - \left[\kappa L\operatorname{sinc}(\xi/2)\right]^2 A^*\left[A^2(t,\tau) \otimes J(\tau)\right] + i\rho A^*$ |

*3.1.2 Effect of non-instantaneous nonlinear response function*

Aiming at understanding the effective third-order nonlinearity and further simplifying the equations in Table 1, we separate Fourier transform of I(τ) and J(τ) into both real [$P(\Omega)$ or $X(\Omega)$] and imaginary [$Q(\Omega)$ or $Y(\Omega)$] parts and individually examine their effects [121,122]. The detailed definition of the nonlinear response functions can be found in Table 2. Here, $P(\Omega)$ [$X(\Omega)$] and $Q(\Omega)$ [$Y(\Omega)$] resemble the dispersive two-photon absorption (TPA) and the dispersive Kerr effect respectively, which can also be interpreted as frequency-dependent nonlinear loss and phase modulation for field at $\omega_0$ during cascaded quadratic process ($\omega_0$, $\omega_0+\Omega \to 2\omega_0+\Omega$). For the doubly resonant cases, the nonlinear response function not only depends on the phase matching condition, similar to the singly resonant case, but also relates to the detuning of field at $2\omega_0$.

**Table 2. Nonlinear response function for singly and doubly resonant cases.**

| Case | Nonlinear response function | Definition |
|---|---|---|
| Singly resonant | $I(\tau) = \mathcal{F}^{-1}[\hat{I}(\Omega)]$ | $\hat{I}(\Omega) = (1 - e^{-ix} - ix)/x^2$, $x(\Omega) = \xi - d_1\Omega - d_2\Omega^2 - i\alpha_2/2$ <br> $\hat{I}(\Omega) = P(\Omega) - iQ(\Omega)$ <br> $P(\Omega) = (1/2)\operatorname{sinc}^2(x/2)$ ($\alpha_2 = 0$) <br> $Q(\Omega) = [1 - \operatorname{sinc}(x)]/x$ ($\alpha_2 = 0$) |
| Doubly resonant | $J(\tau) = \mathcal{F}^{-1}[\hat{J}(\Omega)]$ | $\hat{J}(\Omega) = (\alpha_2 + i\delta_2 - id_1\Omega - id_2\Omega^2)^{-1}$ <br> $\hat{J}(\Omega) = X(\Omega) - iY(\Omega)$ <br> $X(\Omega) = \{\alpha_2[1 + (\Delta_2 - D_1\Omega - D_2\Omega^2)^2]\}^{-1}$ <br> $Y(\Omega) = (\Delta_2 - D_1\Omega - D_2\Omega^2)X(\Omega)$ |

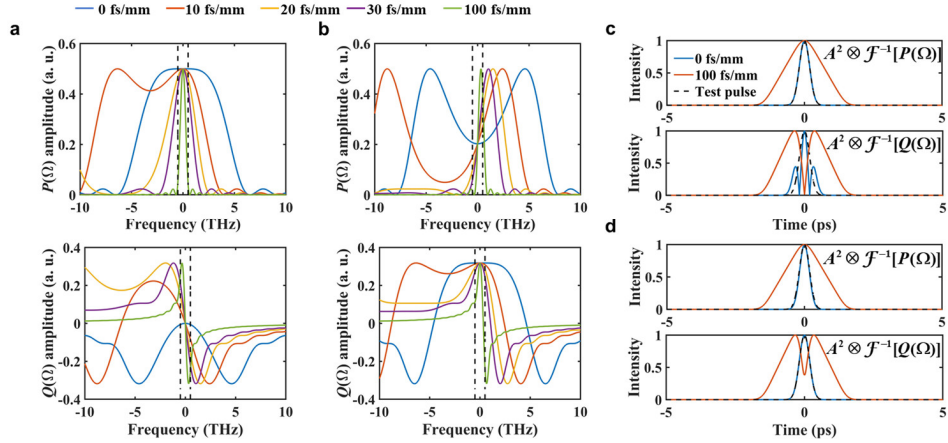

Fig. 11. Walk-off leads to pulse distortion. (a)(b) The real [$P(\Omega)$] and imaginary [$Q(\Omega)$] part of $\hat{I}(\Omega)$ with different GVMs. (c)(d) The convolution of a test pulse with $P(\Omega)$ and $Q(\Omega)$. The dash black lines show the pulse spectral bandwidth in (a)(b) and temporal profiles in (c)(d). (a)(c) perfect phase matching with $\xi=0$; (b)(d) phase mismatch with $\xi=\pi$.

Taking I(τ) as an example, time domain response disappears outside of $0\leq\tau\leq d_1$ for $\Delta k' > 0$ (or $d_1\leq\tau\leq 0$ for $\Delta k' < 0$), since the integrand of I(τ) is an analytic function outside of this region [116]. Therefore, large walk-off results in long duration of non-instantaneous response, which is verified by the narrowband $P(\Omega)$ and $Q(\Omega)$ with large GVM (Fig. 11a and b). However, the delayed response will lead to the strong pulse modulation (Fig. 11c and d), which might cause the pulse break-up and even pulse annihilation, similar to the effect of mode-crossing induced perturbation to the conventional DKS [121,122]. Pulse does not experience severe distortion only when $P(\Omega)$ and $Q(\Omega)$ are approximately flat within the pulse spectral bandwidth, which can be achieved by non-zero phase mismatch and near-zero walk-off. With perfect phase matching condition, $Q(\Omega)$ is always zero at $\Omega = 0$ (Fig. 11a), which means there is no effective Kerr nonlinearity and no cascaded quadratic process due to one-way conversion from pump to either SHG or degenerated signal. Similar conclusions can be drawn for the cases with J(τ).

*3.1.3 Simplified single mean-field equation and bright soliton existing regime*

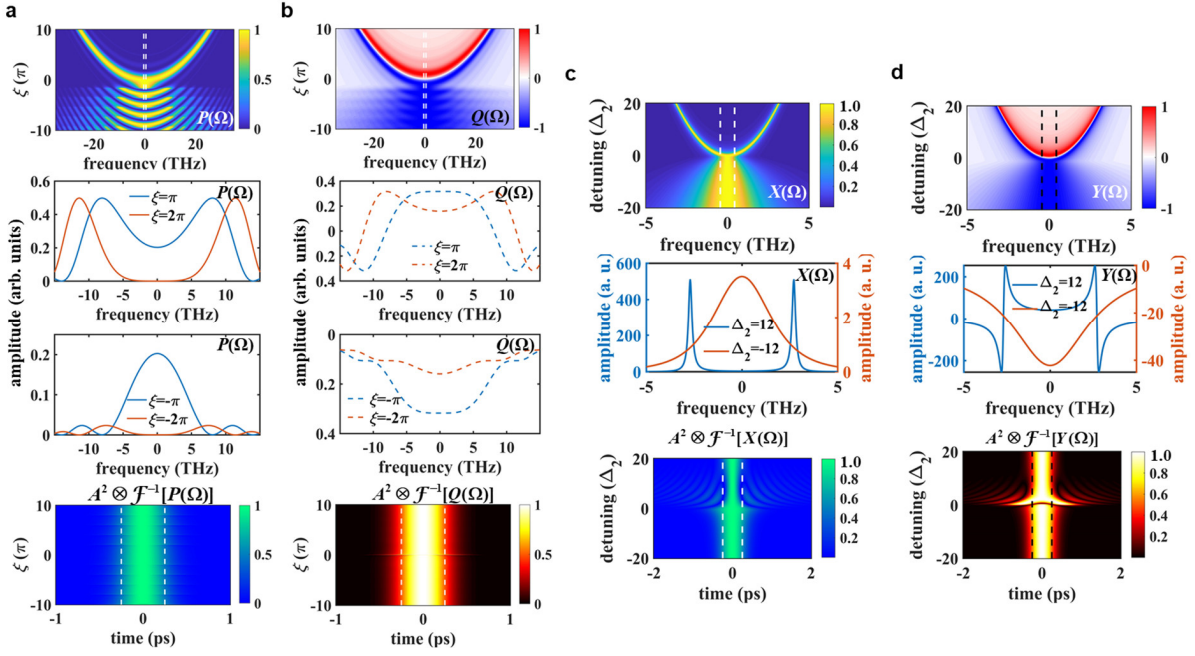

Fig. 12. Frequency response and effect on pulse of dispersive Γ̂(Ω) (a)(b) [122] and Ĵ(Ω) (c)(d) [121]. (a)(b)(c)(d)The first row shows the real and imaginary parts as a function of the wave-vector mismatch parameter. The last row shows the influence on the convolution of the test pulse and inverse Fourier transformation of Γ̂(Ω) and Ĵ(Ω). The middle rows of (a)(b) show line profiles of the real and imaginary parts. Reproduced with permission from Refs. [121,122].

Due to the large walk-off induced soliton perturbation, quadratic soliton is anticipated to exist with small walk-off. Under the condition of zero walk-off, both Γ̂(Ω) and Ĵ(Ω) strongly depends on the other system parameters, including the phase matching condition, the two fields GVDs and detunings, which are summarized in Fig. 12 [121,122]. Two zones, upper zone and lower zone, are discriminated by GVD signs of the two fields (upper zone: different signs, lower zone: same signs). The upper zone exhibits TPA peaks and phase anomalies and causes severe pulse distortion, while the low zone exhibits smooth envelope. This difference leads to different soliton forming mechanisms and corresponding dynamics, which will be discussed in Section 3.2 and 3.4. The bandwidth of Γ̂(Ω) in the upper zone (frequency distance between TPA peaks and phase anomalies) for singly resonant case is characterized by $2\xi/d_2$ (Fig. 12a and b) [122], while for the doubly-resonant case the bandwidth of Ĵ(Ω) is $2\delta_2/d_2$ (Fig. 12c and d) [121]. Usually, phase mismatch parameters ξ is much larger than cavity detuning $\delta_2$ since it requires small $\delta_2$ to lower the threshold. Therefore, it is of advantage to utilize singly resonant structure to alleviate soliton perturbation but at the cost of high pump threshold.

By treating the dispersive Γ̂(Ω) and Ĵ(Ω) as constants at the center frequency, the single mean-field equations in Table 1 can be further simplified to a either LLE-like (for SR-SHG and DR-SHG) or PD-NLSE-like (for SR-DOPO and DR-DOPO) equation as shown in Table 3. Such simplification not only greatly reduces the computational complexity for solving the equations, but also provides more insights about the DQS existence, dynamics, and property. The corresponding parameters can be found in Table 4. The cavity enhanced SHG is described by an equation similar to LLE, while the degenerate OPO is described by an equation similar to PD-NLSE. Compared with LLE or PD-NLSE, there is an additional term representing TPA effect, which evidently increases the threshold. The sign of effective Kerr nonlinearity is mainly determined by the phase matching condition or detuning at $2\omega_0$, which means bright soliton can be achieved with normal dispersion. The existing regime of bright quadratic soliton is summarized in Table 5, which is also confirmed by other simulated results [117].

**Table 3. Simplified single mean-field equation.**

| Case | Single mean-field equation |
|---|---|
| SR-SHG | $t_R \dfrac{\partial A}{\partial t} = -\left(\alpha_1 + i\delta_1 + i\dfrac{k_1'' L}{2}\dfrac{\partial^2}{\partial \tau^2}\right)A - \alpha_{TPA} L|A|^2 A + i\gamma_{eff} L|A|^2 A + \sqrt{\theta_1} A_{in}$ |
| DR-SHG | $t_R \dfrac{\partial A}{\partial t} = -\left(\alpha_1 + i\delta_1 + i\dfrac{k_1'' L}{2}\dfrac{\partial^2}{\partial \tau^2}\right)A - \alpha_{TPA} L|A|^2 A + i\gamma_{eff} L|A|^2 A + \sqrt{\theta_1} A_{in}$ |
| SR-DOPO | $t_R \dfrac{\partial A}{\partial t} = -\left(\alpha_1 + i\delta_1 + i\dfrac{k_1'' L}{2}\dfrac{\partial^2}{\partial \tau^2}\right)A - \alpha_{TPA} L|A|^2 A + i\gamma_{eff} L|A|^2 A + i\mu A^*$ |
| DR-DOPO | $t_R \dfrac{\partial A}{\partial t} = -\left(\alpha_1 + i\delta_1 + i\dfrac{k_1'' L}{2}\dfrac{\partial^2}{\partial \tau^2}\right)A + i\gamma_{eff} L|A|^2 A + i\rho A^*$ |

Table 4. Parameters for Table 3.

| Case | $\alpha_{TPA}$ | $\gamma_{eff}$ | Driving term |
|---|---|---|---|
| SR-SHG | $\kappa^2 L\,\text{sinc}^2(\xi/2)/2$ | $\kappa^2 L[1-\text{sinc}(\xi)]/\xi$ | $\sqrt{\theta_1} A_{in}$ |
| DR-SHG | $\alpha_2 L\dfrac{[\kappa\,\text{sinc}(\xi/2)]^2}{\delta_2^2 + \alpha_2^2}$ | $\delta_2 L\dfrac{[\kappa\,\text{sinc}(\xi/2)]^2}{\delta_2^2 + \alpha_2^2}$ | $\sqrt{\theta_1} A_{in}$ |
| SR-DOPO | $\kappa^2 L\,\text{sinc}^2(\xi/2)/2$ | $\kappa^2 L[1-\text{sinc}(\xi)]/\xi$ | $\mu = \kappa L\,\text{sinc}(\xi/2) e^{-i\xi/2}\sqrt{\theta_p} B_{in}$ |
| DR-DOPO | $0\;(\xi=0)$ | $\dfrac{\kappa^2 L\delta_2}{\delta_2^2 + \alpha_2^2}\;(\xi=0)$ | $\rho = -i\kappa L\sqrt{\theta_2} B_{in}/\sqrt{\delta_2^2+\alpha_2^2}\;(\xi=0,\psi=-\pi/2)$ |

Table 5. Existence of bright quadratic soliton.

| SR-SHG & SR-DOPO | | | | | DR-SHG & DR-DOPO ($\delta_2 = 2\delta_1$) | | | |
|---|---|---|---|---|---|---|---|---|
| soliton regime | $k_1''$ | $k_2''$ | $\xi$ | $\delta_1$ | soliton regime | $k_1''$ | $k_2''$ | $\delta_2$ |
| upper zone | - | + | + | + | upper zone | - | + | + |
| upper zone | + | - | - | - | upper zone | + | - | - |
| lower zone | - | - | + | + | lower zone | - | - | + |
| lower zone | + | + | - | - | lower zone | + | + | - |

## 3.2 Bifurcation analysis and bright quadratic soliton forming mechanism

In this section, we will review the effect of narrowband nonlinear response function on the bifurcation and linear stability of homogeneous states, as well as the soliton forming mechanisms.

*3.2.1 Cavity enhanced SHG*

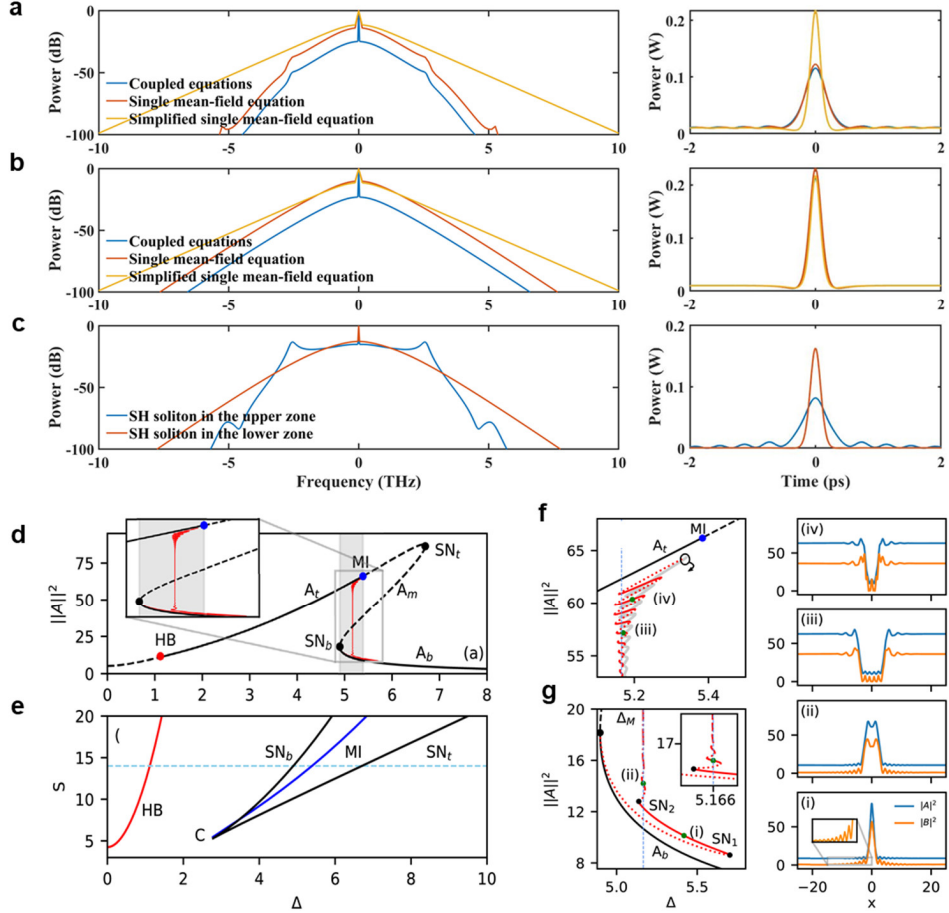

Fig. 13. For DR-SHG case, the simulated quadratic solitons in the upper (lower) zone for FF field in (a) [(b)] by using coupled-wave equations, single mean-field equation and simplified single mean-field equation with same parameters. (c) SH soliton using coupled-wave equation. (d) Cavity resonance where solid (dashed) black line represents stable (unstable) CW state solutions [110]. (e) Main bifurcation instability [110]. (f) Close-up view of the collapsed snaking associated with dark soliton [110]. (g) Collapsed snaking related to the bright soliton [110]. Reproduced with permission from Ref. [110].

For DR-SHG cases, by comparing the simulated results of coupled-wave equations (full equations), equation in Table 1, Table 3 (Fig. 13a, b and 13c), it is found that (i) equation in Table 1 has a good approximation to the couple-wave equations and can repeat almost identical results; (ii) equation in Table 3 is over-approximated especially for the upper zone due to the neglected the TPA peaks and phase anomalies, which invade the soliton existing regime and change the soliton characteristics. According to the bifurcation and linear stability analysis of the coupled-wave equations (Fig. 13d) [110], quadratic soliton in the upper zone is formed by locking of domain walls connecting two coexisting CW states, instead of the balance between the nonlinearity and dispersion for conventional DKS. As for the upper branch of the titled resonance curve, there exists a stable regime through Hopf bifurcation before MI occurs. Therefore, it shows bistability in the detuning regime between saddle node (SN) and where MI starts. In this regime, both the upper branch and lower branch are stable and can coexist, which is totally different from the conventional LLE with unstable upper branch and stable lower branch [124]. Due to the limit of space, the bifurcation and linear stability analysis in the lower zone as well as the SR-SHG case, will not be presented here, since they can be easily accessed with the same picture of DKS governed by LLE [124] owing to the broadband nonlinear response function.

*3.2.2 Degenerate OPO*

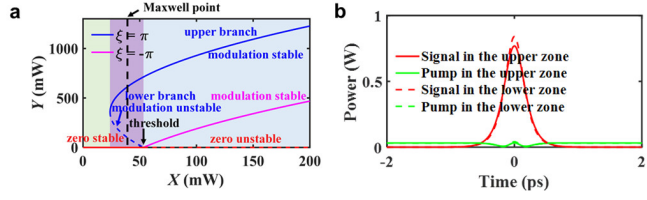

Fig. 14. SR-DOPO case. (a) Bifurcation diagram for the steady-state solutions including subcritical bifurcation (blue lines) and supercritical bifurcation (magenta lines) [122]. (b) Simulated signal and pump pulse profiles using coupled-wave equations. Reproduced with permission from Ref. [122].

We will start from the SR-DOPO case and then switch to the DR-DOPO one. As for SR-DOPO cases, the broadband nonlinear response function, either in the upper zone or the lower zone, ensures the correctness of the further simplified single mean-field equation in Table 3. As shown in Fig. 14a, the bifurcation and linear stability (using equation in Table 3) [122] display a bistable regime below threshold for $\xi = \pi$, where both the zero solution and upper branch are stable while the lower branch is modulationally unstable. Signal soliton is formed through locking of fronts connecting the two stable solutions (domain wall locking): the stable zero solution and stable upper branch solution, resulting in "*topological soliton*" without ripples in the soliton tails. Owing to the discrete phase symmetry $A \rightarrow -A$, the bright soliton exhibit two phase states (for the upper branch) with opposite phase, which adds a degree of freedom compared with conventional DKS and benefits applications such as random bit generation [107] and Ising machines [108].

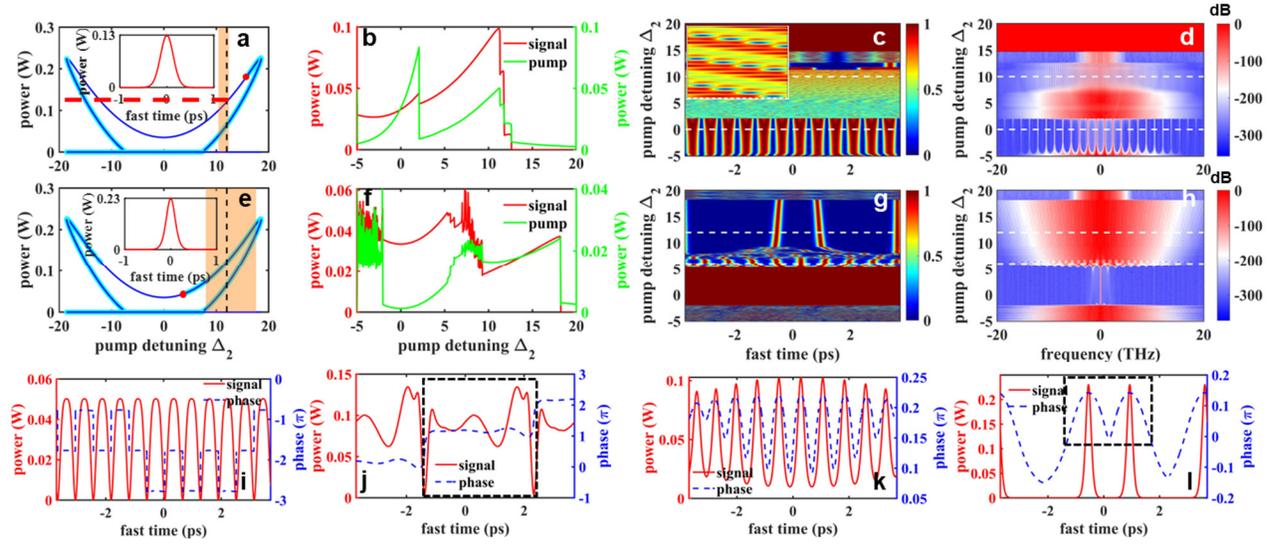

Fig. 15. Dynamics of DR-DOPO via frequency scanning [125]. Upper zone: (a)–(d), (i) and (j). Lower zone: (e)–(h), (k) and (l). (a)(e) resonance diagrams and linear stability analysis for the non-zero solutions and zero solutions. The light orange shaded areas indicate the soliton existing regime. (b)(f) intra-cavity average power evolution, (c)(g) signal pulse evolution and (d)(h) signal spectrum evolution with $\Delta_2$. (i)(j) signal profiles and corresponding phase as snapshots at $\Delta_2 = 0$ and $\Delta_2 = 10$. The dashed box in (j) shows an in-phase localized structure. (k)(l) signal profiles and corresponding phase as snapshots at $\Delta_2 = 6$ and $\Delta_2 = 12$. The dashed boxes in (k) show in-phase localized structures.

As for DR-DOPO in the upper zone, quadratic soliton obtained by equation in Table 3 is overestimated, similar to the DR-SHG case. The bifurcation and linear stability results in Fig. 15a [125] indicate the quadratic soliton in the upper zone is referred to the *topological soliton* which is formed by domain wall locking between the stable upper branch and stable zero solution, verified by the identical power between soliton peak power (inset of Fig. 15a) and the CW power of upper branch. On the contrary, the quadratic soliton in the lower zone is a conventional soliton due to the balance of GVD and nonlinearity with unstable upper branch and stable zero solution in the soliton existing regime.

*3.3 Dark quadratic soliton*

Dark soliton in conventional Kerr cavities is formed through domain wall locking between the stable upper branch and stable lower branch [122]. The minimum power of the dark Kerr soliton is not zero, but close to the power of the lower branch. Here dark quadratic soliton can also be formed through this mechanism, which can be realized for cases of SR-SHG and DR-SHG in both upper and lower zone, with governing equations similar to LLE. Besides, dark quadratic soliton can also be formed in degenerate OPOs by domain wall locking between two stable states with opposite phases, which is the intrinsic nature of OPO owing to the discrete phase symmetry $A \rightarrow -A$. For SR-DOPO case, dark quadratic soliton (also termed as Ising wall), must be generated in pairs to satisfy cavity boundary condition (Fig. 16). The peak power of the stable dark soliton is identical with the upper branch, while the minimum power is zero. The same forming mechanism of dark quadratic soliton can be applied to DR-DOPO cases.

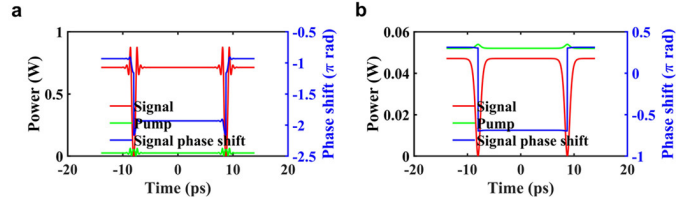

Fig. 16. SR-DOPO case. Dark quadratic soliton formed through domain wall locking between two stable states with opposite phases [122]. Simulated dark soliton pairs in the blue area of Fig. 14a for the blue upper branch (a) and magenta upper branch (b). Reproduced with permission from Ref. [122].

*3.4 Quadratic soliton dynamics with pump frequency scanning*

The narrowband nonlinear response function especially in the upper zone for doubly resonant cases, will strongly influence the bifurcation and linear stability, resulting in different dynamics. Fig. 15 shows the example of DR-DOPO dynamics with pump frequency scanning.

In the upper zone, during the tuning process localized structures (Fig. 15i) start from initial noise through domain wall locking between two stable upper branches out of phase, leading to the formation of Ising walls. As the detuning continues going to the red side, the Ising walls transit to nonstationary Bloch walls with a constant drifting velocity (the inset of Fig. 15c) owing to the chiral symmetry breaking. This Ising-Bloch transition [126] triggered new localized structures and grey pulses as shown in Fig. 15j. When the detuning goes deeply into the soliton existing regime, the grey pulses cannot be kept stable and *topological solitons* form as another kind of domain wall, which connects a stable upper branch and a stable zero solution. According to Fig. 15b, evident soliton step is formed during the frequency tuning process, similar to conventional DKS.

In the lower zone, when the pump frequency is tuned from blue side to red side, the system experiences CW, MI and then goes into the soliton existing regime. Soliton is shaped from the MI pulse by the balance between GVD and nonlinearity. First, CW state or domain wall locked structures are triggered from initial noise. When the detuning continues going to red side, Turing patterns (Fig. 15k) then occurs through MI from the unstable CW state or domain wall locked structures. When the detuning goes deeply into the soliton existing regime, the in-phase patterns close to each other will collapse and form solitons or disappear, while the out-of-phase patterns will repulse each other and evolve into solitons (Fig. 15l).

Of note, the soliton existing regime in Fig. 15e is much broader than that in Fig. 15a. The reason can be: (i) soliton perturbation resulting from the TPA peaks and phase anomalies in the upper zone. (ii) for *topological soliton* formation, the group velocity difference (v) of the two domain walls depends on the detuning and should be limited in a small value around the Maxwell point [112].

*3.5 Quadratic soliton in non-degenerate OPOs*

By simplifying the three-wave coupled equations into two-coupled PD-NLSEs, symbiotic solitons trapped with each other in doubly resonant non-degenerate OPOs are found with zero walk-off between the interacting three waves (Fig. 17a). Similar to cross-phase induced trapped DKS, symbiotic solitons with sech$^2$ profile as well as with flat-top profile (Fig. 17b) can be generated with same or opposite signs of GVD for signal and idler fields [123].

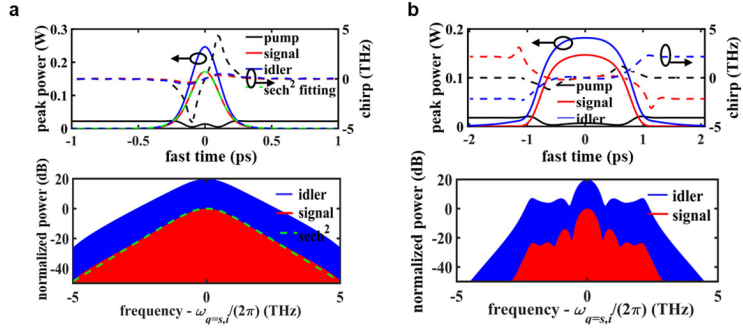

Fig. 17. Quadratic solitons in non-degenerate OPOs. (a) symbiotic solitons with same signs for both signal and idler. (b) symbiotic solitons with opposite signs for signal and idler. Reproduced with permission from Ref. [123].

*3.6 Quadratic soliton perturbation from non-zero walk-off*

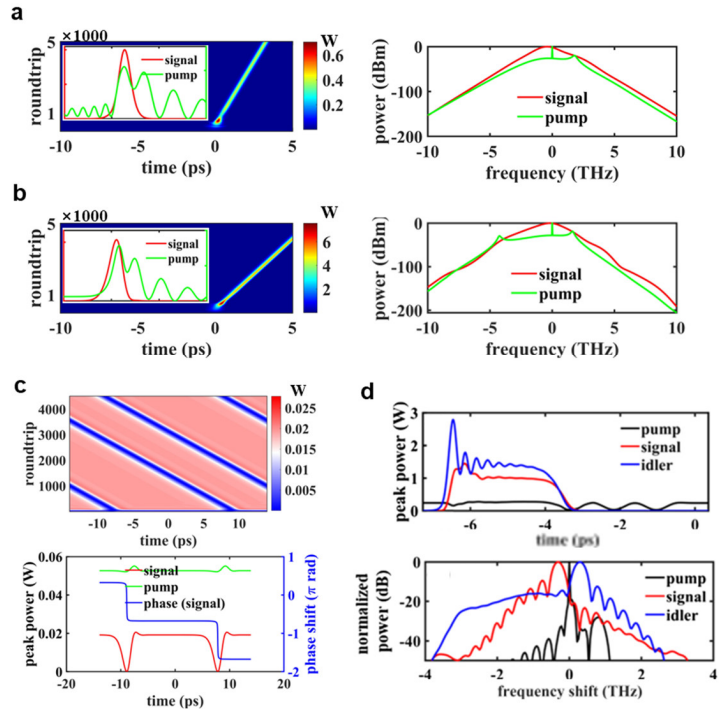

Fig. 18. The effect of walk-off on quadratic solitons. Bright soliton in DR-DOPO for the upper branch (a) and lower branch (b) [121]. Dark soliton pairs in SR-DOPO (c) [122] and non-degenerate OPO (d) [123]. Reproduced with permission from Refs. [[121], [122], [123]].

With non-zero and small walk-off, the soliton profile deviates from a symmetric $sech^2$ profile, including a sharp tail on one side (Fig. 18) [121]. In this condition, walk-off affects the soliton in a way similar to the third-order dispersion on DKS [127]. The maximum walk-off that the solitons keep stable (tolerant walk-off) is related to not only the system parameters themselves, but also the resonant scheme. Despite the advantage of low threshold for DR-DOPO, it is more sensitive to the effect of temporal walk-off. In general, doubly-resonant cases have less tolerance (~10 fs) to the walk-off than the singly-resonant cases (~1 ps). This can be intuitively understood by the fact that it requires small walk-off to keep the two resonant solitons trapping with each other [84]. Walk-off effect is dramatically weakened for singly-resonant cases because the one of the fields escapes the cavity in time.

With large walk-off, quadratic solitons are severely perturbed by the TPA peaks and phase anomalies near the center frequency, thus evolving into the regime dominated by the walk-off induced MI, as shown in Ref. [103]. Self-organized phase-locked OFCs can be generated due to the balance of nonlinear gain and loss, nonlinearity and dispersion, and the boundary conditions, although without fully stable temporal pattern.

*3.7 Summary and experimental guidelines*

In summary, (i) pure quadratic soliton exists with a small walk-off between two fields; singly resonant configurations result in larger walk-off tolerance than doubly resonant ones at the cost of higher pumping power; (ii) *topological soliton* exists in the upper zone with opposite GVD signs for the two fields, while conventional soliton exists in the lower zone with same GVD signs; (iii) existing regime for *topological soliton* is much narrower than the conventional soliton. Therefore, to experimentally achieve pure quadratic soliton, careful dispersion engineering should be conducted for near zero-GVM and small GVD value with same signs.

**4. Cavity soliton generation with coexisting Kerr and quadratic nonlinearities**

Section [3](#) mainly focuses on the pure quadratic soliton in $\chi^{(2)}$ cavities without considering $\chi^{(3)}$ nonlinearity. In fact, Kerr nonlinearity coexists with the quadratic nonlinearity and may compete with other one, which will trigger many novel phenomena in nonlinear optics. For example, the recent demonstration of octave-spanning spectrum generation on a nanophotonic PPLN waveguide [128], proves the high efficiency of spectral broadening with the interacting $\chi^{(2)}$ and $\chi^{(3)}$ nonlinearity. Another example is the widely tunable femtosecond Kerr soliton generation achieved in a synchronously pumped nondegenerate OPO by inserting a long fiber (Kerr media) in a free-space cavity [129]. Here we will review soliton generation in microresonators with coexisting Kerr and quadratic nonlinearities.

*4.1 Cavity soliton formation with dominant quadratic or Kerr nonlinearity*

*4.1.1 DKS formation with dominant quadratic nonlinearity*

Material Kerr nonlinearity can cause perturbation to the quadratic soliton when it is comparable to the effective Kerr nonlinearity resulting from the cascaded quadratic process. Besides, soliton dynamics and existing regime can be altered due to the competing nonlinearities [130].

Quartic nonlinearity can be utilized to flip the sign of effective Kerr nonlinearity and re-locate the soliton existing regime in the thermally stable, blue-detuned regime. For most of the materials with positive thermo-optic coefficient and positive Kerr nonlinearity, DKS generation is prohibited in the unstable red-detuned regime due to the thermal nonlinearity. In order to engineer soliton existing regime to be thermally stable, signs of either thermo-optic coefficient or the Kerr nonlinearity can be altered. The PR effect mentioned in Section [2](#) is a good example since it is a stronger effect than the thermo-optic effect with opposite sign. Another example is to alter the sign of total Kerr nonlinearity through the cascaded quadratic process [131]. By exploiting the opposing interactions of a slow thermal response and a fast, negative, Kerr nonlinearity in a periodically poled lithium niobate (PPLN) microresonator with normal dispersion, a novel operating regime is theoretically found where multistability is broken and deterministic single DKS access is guaranteed. The underlying CW-only solution is unstable, while the single DKS is the only stable behavior, termed as monostable DKS. The three essential criteria that define MS-DKS are (i) to break the multistability of solitons, (ii) to open a monostable window, and (iii) to position the single soliton existence range within the monostable window. We observe cycling through behaviors such as chaos, soliton states, and CW-only that settles into stable single soliton behavior, indicating self-starting, deterministic and robust operation.

*4.1.2 DKS formation with dominant Kerr nonlinearity*

As discussed in Section [2](#), bright DKS can be perturbed by linear and nonlinear effects, while dark soliton can be initiated through MI induced by these effects. Similar phenomena are theoretically predicted due to the perturbation from the $\chi^{(2)}$ process induced localized anomalies of loss and phase, including the soliton number reduction and single bright soliton generation [[132], [133], [134]]. Soliton crystal generation is observed due to the soliton interaction resulting from nonlinear mode coupling induced background modulation [135]. The weak nonlinear mode coupling through $\chi^{(2)}$ process can also lead to dark soliton generation in microresonators with normal dispersion. Assisted by the weak quadratic nonlinearity in $Si_3N_4$ microresonators, the generated narrowband second harmonic field opens a channel to trigger the MI and eventual dark soliton comb for the pump field (Fig. 19) [136].

The coexisting nonlinearities also trigger dual-color soliton in $LiNbO_3$ microresonator. We owe the soliton at the second harmonic in Ref. [76] into a transferred soliton from DKS at the fundamental field. In this case, the imperfect phase matching condition leads to weak cascaded quadratic process thus weak effective Kerr nonlinearity compared to the material Kerr nonlinearity.

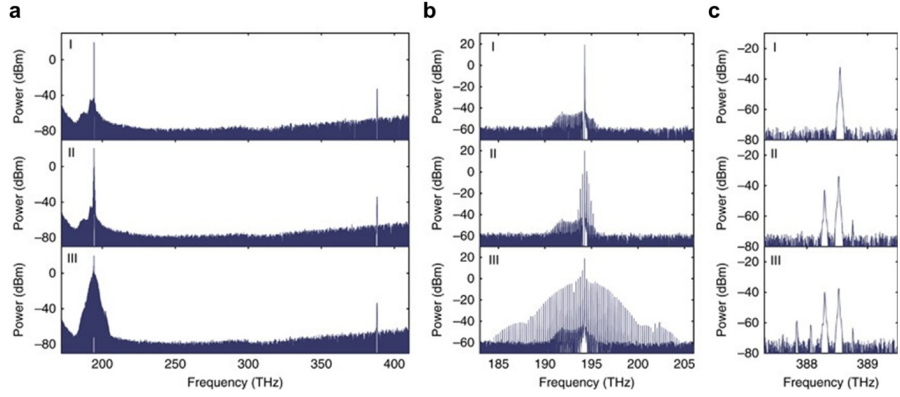

Fig. 19. Second-harmonic generation assisted dark soliton generation. (a) spectra measured in a range of more than one octave. (b) zoomed-in infrared spectra. (c) zoomed-in second-harmonic spectra. Reproduced with permission from Ref. [136].

### *4.2 Parametrically driven DKS (soliton formation with dominant quadratic and Kerr nonlinearity)*

Parametrically driven dissipative Kerr soliton (PD-DKS) refers to soliton in OPO where its pulse energy is sustained by the pump through quadratic nonlinearity, but its pulse shape is defined by the balance between GVD and Kerr nonlinearity. This is in stark contrast to the DQS described in Section 3, where the cascaded quadratic nonlinearity is also responsible in providing the effective Kerr nonlinearity for soliton pulse shaping. PD-DKS generation in the synchronously pumped all-fiber resonator is shown in Ref. [137]. The quadratic nonlinearity is provided by a short periodically poled fiber while the Kerr nonlinearity is provided by a long passive fiber. A segment of erbium-doped fiber is also inserted in the cavity to compensate the intracavity loss, enhance the quality factor, and consequently boost the Kerr phase shift. To avoid soliton perturbation, lasing is suppressed by employing a bandpass filter and operating below the threshold. Importantly, similar to the results in Ref. [67] where DKS is generated in an active fiber cavity, the gain is also in the weak saturation regime so that PD-DKS can be triggered from MI. However, different from conventional DKS in Ref. [67], here PD-DKS with center wavelength of ~1550 nm is realized by pumping at ~775 nm through the degenerate parametric process in the periodically poled fiber. In addition, the non-resonant pump laser emits from the cavity in time such that the walk-off between the pump and signal does not severely influence the soliton generation. The bifurcation diagram suggests that the upper branch is modulationally unstable. Seeded from the MI in the upper branch, the PD-DKS here is generated through the balance of GVD and Kerr nonlinearity. By scanning the pump frequency, a clear soliton step is observed as well as the chaos before the soliton generation.

This first PD-DKS demonstration in the all-fiber resonator suffers high threshold resulting from singly resonant configuration, low-Q and weak quadratic nonlinearity of the periodically poled fiber. Synchronous pumping that inevitably adds system complexity is thus required. Recent demonstration of PD-DKS in a monolithic high-Q AlN microresonator with CW pump solves the above-mentioned problems [106]. The doubly resonant configuration as well as the high-Q efficiently reduce the soliton threshold to ~100 mW. The much stronger quadratic nonlinearity greatly enhances the power conversion efficiency to 17% from pump at ~775 nm to background-free soliton at ~1550 nm. *Our group* theoretically study the nature of PD-DKS in a DR-DμOPO [138] and show that there is a threshold GVM above which single PD-DKS in DR-DμOPO can be generated deterministically. Our simulation is in a good agreement with the experimental results shown in Ref. [106] (Fig. 20). Importantly, the perturbative effective Kerr nonlinearity resulting from the GVM-dependent cascaded quadratic process is responsible for the soliton annihilation and the deterministic single PD-DKS generation. Finally, we provide the design guidelines for accessing such deterministic single PD-DKS state.

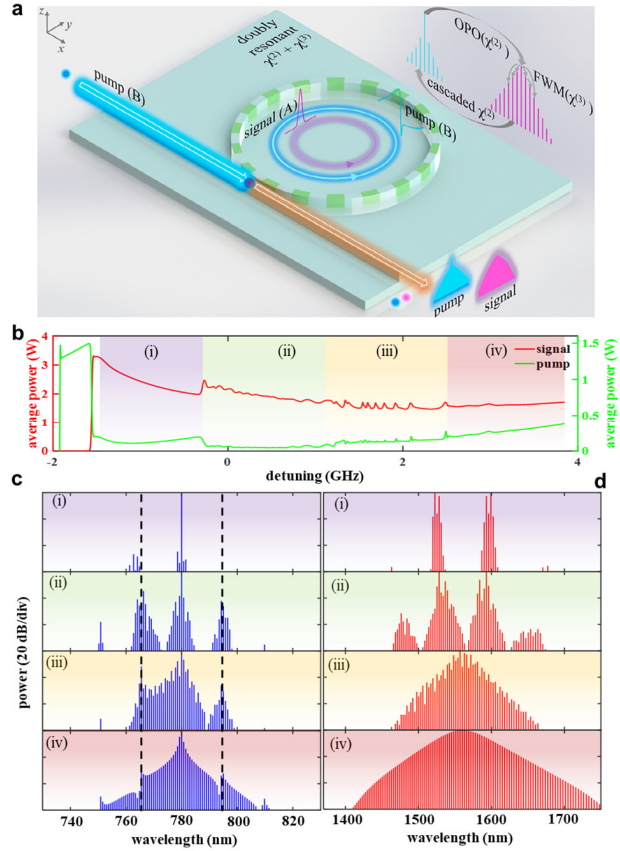

Fig. 20. Simulated PD-DKS dynamics in AlN microresonator [138], which agrees well with Ref. [106]. Reproduced with permission from Ref. [138].

## 5. Metrology tools for photonic frequency microcomb

### 5.1 Ultrafast DKS dynamics

Real-time experimental observation of DKS dynamics, such as soliton generation, collision, and annihilation, is challenging because they exhibit transient and non-repetitive temporal features on sub-ps time scale and evolve over nanosecond to microsecond, which is beyond the capability of any conventional measurement approaches. Developing innovative measurement tools that provide ultrahigh detection bandwidth and long record length in single-shot and real-time manner is thus critical for the deeper understanding of DKS frequency comb. Here we review the emerging ultrafast measurement tools that have been developed for the study of real-time DKS dynamics.

#### 5.1.1 Electro-optic (EO) comb sampling

The principle of EO comb sampling method is illustrated in the Fig. 21. A EO frequency comb is generated by EO-modulating a cCW single frequency laser at a modulation frequency $f_M$ close to the FSR of the microresonator using a combination of intensity and phase modulators. The EO comb and the DKS comb are then combined together and detected by a photodetector (PD), which generates an interferogram shown in the lower panel. Owing to the repetition rate difference ($\Delta f_{rep}$) between the EO comb and the DKS comb, the EO comb pulse train gradually walks off from the DKS pulse train and performs asynchronous optical sampling where the envelope of the interferogram represents the temporally magnified DKS waveform. The measurement frame rate is equal to $\Delta f_{rep}$, and the measurement temporal resolution is determined by the EO-comb pulse width, which reaches sub-ps level limited by the modulation depth and bandwidth of the EO modulators.

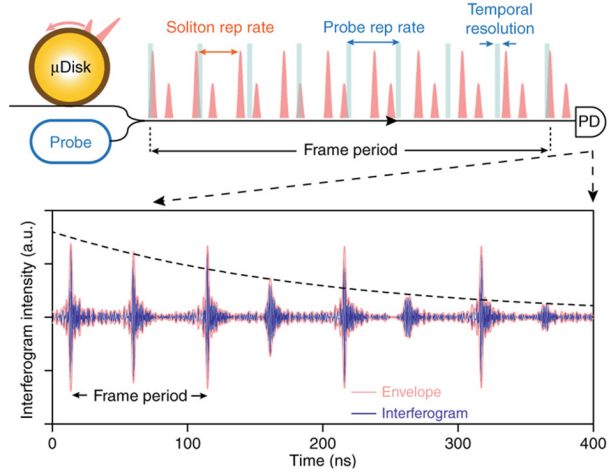

Fig. 21. (a) Schematic diagram of EO comb sampling. (b) Example of coherent sampling waveform. Reproduced with permission from Ref. [89].

The first real-time measurement of DKS dynamics using EO comb sampling was demonstrated in 2018 [89]. With a measurement temporal resolution of 800 fs and a measurement frame rate of 10 MHz, the complete single DKS generation process during 50-μs pump detuning sweep was observed. Soliton collision as well as the temporal and spectral breathing of breather soliton have also been observed. Very recently, EO comb sampling technique has been utilized to resolve the details of heteronuclear soliton molecules in optical microresonators generated by strongly phase-modulated pump [90], as well as the soliton formation and collision in multistability regime generated by bichromatic pump [88].

However, much like the dual-comb spectroscopy, EO-comb sampling is not a single-shot measurement technique due to the asynchronous sampling nature. Each measurement is averaged over $N$ roundtrips expressed by

$$N = \frac{2}{\tau f_M}, \tag{2}$$

where $\tau$ is the measurement temporal resolution. For the system demonstrated in Ref. [88], a maximum measurement frame rate of 200 MHz can be achieved at 1-ps resolution for a 22-GHz FSR DKS frequency comb, which corresponds to an averaging of 110 roundtrips despite the already high measurement frame rate. Therefore, EO comb sampling is more suitable for measuring slowly evolved dynamics that is quasi-static over $N$ roundtrips.

*5.1.2 Temporal magnifier*

Another powerful measurement technique that has been utilized to study the DKS dynamics is the ultrafast temporal magnifier [139,140] that is based on the space-time duality principle [141,142] (Fig. 22). In contrast to EO-comb sampling technique, the temporal magnifier is fundamentally a single-shot measurement technique that can thus resolve DKS dynamics roundtrip by roundtrip. Much like a single-lens imaging system generating a spatially magnified image from a small object, the temporal magnifier consists of two chromatically dispersive element such as optical fibers to provide input group-delay dispersion (GDD) Φ1″ and output GDDΦ2″, and a time-lens applying a quadratic temporal phase modulation as

$$\phi(t) = \frac{t^2}{2\Phi_f''}, \tag{3}$$

where $\Phi_f''$ is the focal GDD of the time lens. When $\Phi_1''$, $\Phi_2''$ and $\Phi_f''$ satisfy a "temporal imaging" condition written as

$$\frac{1}{\Phi_1''} + \frac{1}{\Phi_2''} = \frac{1}{\Phi_f''}, \tag{4}$$

the DKS waveform under test will be temporally magnified by a factor of

$$M = \left|\frac{\Phi_2''}{\Phi_1''}\right|. \tag{5}$$

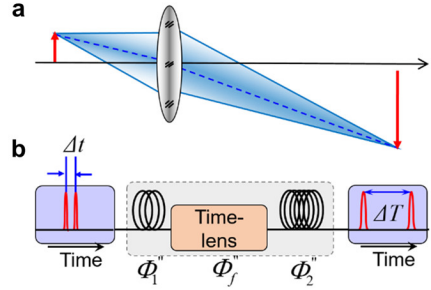

Fig. 22. Principle of temporal magnifier. (a) Spatial analog of time magnifier. (b) Schematic diagram of a time magnifier. Reproduced with permission from Ref. [142].

State-of-the-art temporal magnifiers use parametric processes such as four-wave mixing to implement the time lens. Its measurement frame rate and temporal resolution, determined by the parametric pump repetition rate and the inverse of the parametric pump bandwidth respectively, are compatible if not better than those of the EO-comb sampling technique. *Our group* has utilized temporal magnifier for real-time characterization of 88-GHz FSR DKS frequency comb dynamics in both uniform and dispersion managed microresonators (Fig. 23) [139]. With a measurement temporal resolution of 600 fs and a measurement frame rate of 250 MHz, ultrafast DKS transitions during pump detuning sweep can be clearly resolved and such study experimentally proved for the first time that dispersion managed microresonators provide a single-soliton stability zone an order of magnitude larger than conventional uniform microresonators.

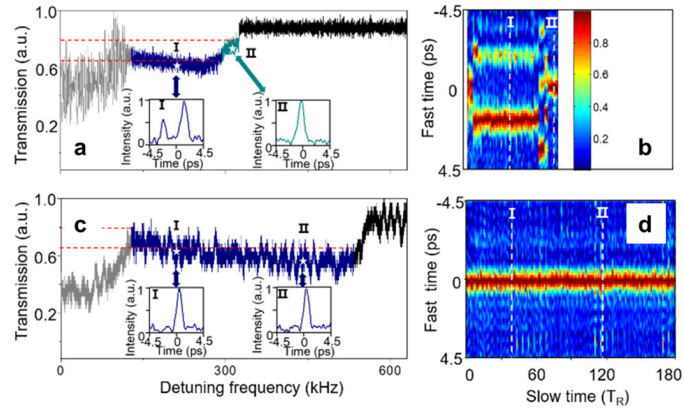

Fig. 23. Comparison of stability zone (a) (c) and evolution dynamics (b),(d) between static (a)(b) and dispersion managed (c)(d) Kerr solitons. Reproduced with permission from Ref. [139].

Just like there is always a field-of-view limitation in any spatial imaging systems, the temporal record length of continuous single-shot measurement using conventional temporal magnifier has been limited to hundreds of picoseconds by the parametric pump bandwidth and the focal GDD of the time lens. To overcome this limitation, *our group* invented the panoramic reconstruction temporal imaging (PARTI) technique that integrates the ultrafast temporal magnifier and the mosaic waveform stitching to combines the feats of femtosecond temporal resolution and nanosecond record length [139]. PARTI is thus suitable for the comprehensive depiction of DKS transient phenomena and non-equilibrium behaviors.

Fig. 24 shows the working principle of the PARTI system. The optical buffer generates multiple high-fidelity replicas (represented by blue, green and red, respectively) of the signal under test (SUT) and the subsequent temporal magnifier captures different portions of the SUT waveform on each replica. After data processing on the system output, the original long SUT waveform is reconstructed through waveform stitching. With 10 optical buffers, the proof-of-principle PARTI has achieved a continuous single-shot measurement over 1.5 ns and revealed ultrafast DKS evolution roundtrip by roundtrip for the first time. With further optimization, high-fidelity waveform stitching of 100 optical buffers should be achievable to scale up the temporal record length to 30 ns and facilitate the study of DKS formation pathways, destabilization routes, and responses to environmental perturbations.

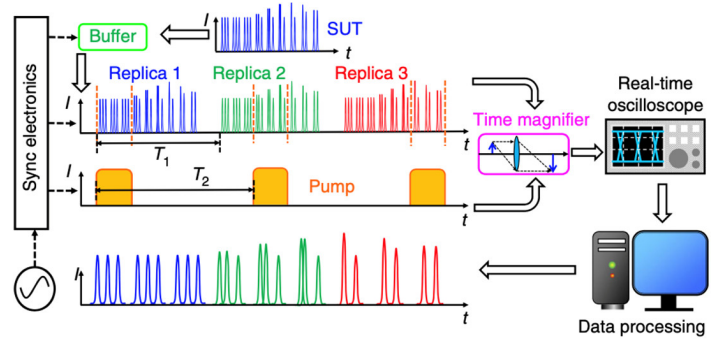

Fig. 24. Principle of the PARTI system. Reproduced with permission from Ref. [140].

*5.2 Narrow comb linewidth*

A common method to measure the fundamental comb linewidth is the short-delayed self-heterodyne interferometer [53,[143], [144], [145], [146]] that characterizes the white frequency noise floor of the comb lines. A phase noise analyzer is usually used to measure the phase noise power spectral density (PSD) of the self-heterodyne signal, but cross-correlation of simultaneously detected self-heterodyne signals can also be implemented to further suppress the photodetection technical noise that limits the measurement sensitivity especially at high offset frequencies [48,147]. Single-sideband (SSB) phase noise PSD $L_\phi(f)$ in the unit of dBc/Hz is then converted to SSB frequency noise PSD $L_\nu(f)$ in the unit of Hz$^2$/Hz according to the equation

$$L_\nu(f) = \frac{f^2}{4\sin^2(\pi f \tau)} L_\phi(f), \tag{6}$$

where $\tau$ is the interferometer delay and $f$ is the offset frequency. Finally, the fundamental comb linewidth $\Delta\nu$ in the unit of Hz is calculated from the white frequency noise floor $L_{\nu w}$ according to the equation

$$\Delta\nu = \pi L_{\nu w}. \tag{7}$$

*5.3 Ultralow DKS timing jitter (phase noise of comb repetition rate)*

While the phase noise of comb repetition rate can be conveniently measured by phase detector methods, the measurement noise floor is typically limited to −160 dBc/Hz by photodetection shot noise and electronics thermal noise [55]. This noise floor is too high for timing jitter characterization of quantum-limited DKS, especially at high offset frequencies. Besides, system complexity of phase noise measurement significantly increases for comb repetition rate larger than 40 GHz. To enhance the measurement sensitivity, BOC was developed and −200 dBc/Hz SSB phase noise floor for 10-GHz carrier has been demonstrated. BOC achieves such unsurpassed sensitivity by taking advantage of another low-noise comb as the reference oscillator and implementing direct optical pulse-to-pulse timing comparison using nonlinear crystals. Details about the BOC design can be referred to Refs. [148,149]. In the microcomb community, BOC has been applied to the study of relative timing jitter between the counter-propagating DKSs and the co-propagating DKS pair [87].

The major drawback of the BOC is the system complexity especially the need for a low-noise comb reference oscillator. In addition, high-peak-power pulses are required to efficiently drive the nonlinear optical process, rendering BOC less suitable for high-repetition-rate DKS measurement. Thus, a timing jitter measurement method that does not require comb reference oscillator and nonlinear optical process is desirable for simple yet accurate characterization of the DKS frequency comb. All-fiber reference-free Michelson interferometer (ARMI) timing jitter measurement setup was developed to answer such a need, and −180 dBc/Hz SSB phase noise floor for 10-GHz carrier has been demonstrated [59]. ARMI has been applied to the study of quantum-limited DKS timing jitters in both fiber Fabry-Perot resonator [55] by *our group* and on-chip silica platform by others [150].

Fig. 25 shows the schematic of the ARMI timing jitter measurement setup [59]. Two well-separated spectral bands ($f_m = mf_{rep} + f_{ceo}$ and $f_n = nf_{rep} + f_{ceo}$) are first filtered out and injected into an unbalanced Michelson interferometer (UMI) for converting phase noise to amplitude noise. The spectral separation is chosen to optimize the detection sensitivity by striking a balance between the UMI scale factor and the filtered comb power. Similarly, the length of the UMI fiber delay line is chosen to optimize the compromise between the detection sensitivity proportional to the delay time $\tau$ and the measurement bandwidth scaling with $1/\tau$. A delay control unit consisting of two motorized delay lines is implemented to compensate for the dispersion of the fiber delay line such that the interferometer output signals for both spectral bands are simultaneously maximized. An acousto-optic frequency shifter (AOFS) with frequency shift of $f_{AOM}$ is included to frequency upshift the interferometer output signal by $2f_{AOM}$ to avoid the electronics noise in the baseband. Finally, the two interferometer output signals are mixed in a double balanced mixer to reject the common-mode $f_{ceo}$ noise. The amplitude noise PSD of the mixer output $L_A(f)$ now contains the information of the comb repetition rate frequency noise $L_\nu(f)$ with the conversion equation

$$L_A(f) = \left[(m-n)\tau\right]^2 V_{pp}^2 \frac{\left|1-e^{-i2\pi f\tau}\right|^2}{4f^2} L_\nu(f), \tag{8}$$

where $V_{pp}$ is the peak-to-peak mixer output voltage. Timing jitter and phase noise can then be calculated from the frequency noise PSD. Of note, a delay-locked loop (DLL) acting on a PZT stretcher is engaged to stabilize the UMI fiber delay line to the comb under test and consequently avoid the bias drift in the UMI during the measurement. It is important to keep the DLL bandwidth as low as possible because the timing jitter measurement is only valid beyond the DLL bandwidth.

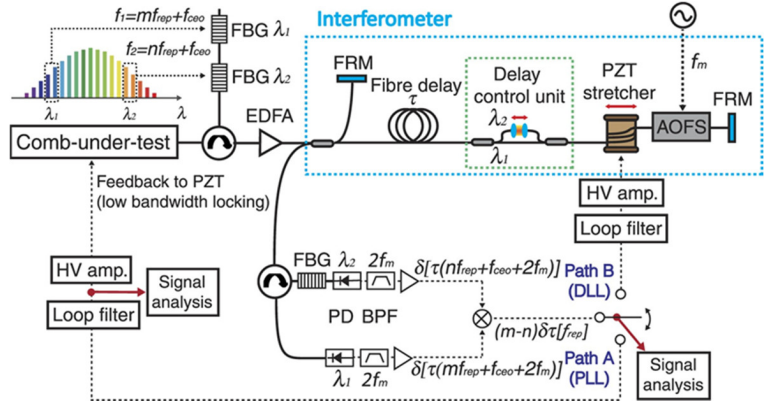

Fig. 25. ARMI setup for comb timing jitter measurement. Reproduced with permission from Ref. [59].

## 6. Application and Outlook

More than a decade of extensive research has led to the maturation of photonic frequency microcomb technology and a plethora of applications with microcombs have been demonstrated recently. Photonic frequency microcomb has rapidly approached the performances of traditional laser frequency comb and become a competitive light source architecture for comb enabling research fields such as optical atomic clockwork and optical frequency synthesizer [[151], [152], [153], [154], [155], [156], [157]]. State-of-the-art microcomb clockwork has achieved a $10^{-17}$ fractional-frequency accuracy and precision for a continuous and glitch-free operation of 2 h [154]. State-of-the-art microcomb synthesizer has achieved a $7.7 \times 10^{-15}$ fractional-frequency uncertainty with the optical frequency programmable across the telecommunications C-band at a 1-Hz resolution [158].

Photonic frequency microcomb has also shown promises in applications including precision spectroscopy [[158], [159], [160], [161]], astrospectrograph calibration [162,163], biomedical imaging [[164], [165], [166]], optical communications [[167], [168], [169], [170], [171], [172], [173]], and coherent ranging [[174], [175], [176]]. In particular, photonic frequency microcomb's large comb spacing renders it outperforms traditional laser frequency comb for optical communications in data rate and coherent ranging in acquisition speed. State-of-the-art microcomb optical communications system can transmit a data stream of more than 50 Tbits/s over 75 km with 179 channels that span the entire telecommunication C and L bands [168]. Coherent ranging with either dual-comb [174,175] or frequency-modulated continuous-wave (FWCM) [176] approach has been demonstrated. State-of-the-art dual-microcomb ranging has achieved a 12-nm distance precision with a 100-MHz acquisition speed at a 13-μs averaging time, allowing for in-flight sampling of gun projectiles moving transversely at 150 m/s or Mach 0.47 (Fig. 26). In the

FMCW microcomb ranging, 30 FMCW channels are utilized to achieve parallel distance and velocity measurements at an equivalent rate of 3 Mpixels/s with the potential to improve beyond 150 Mpixels/s, two orders of magnitude higher than the traditional eye-safe FMCW LiDAR [176].

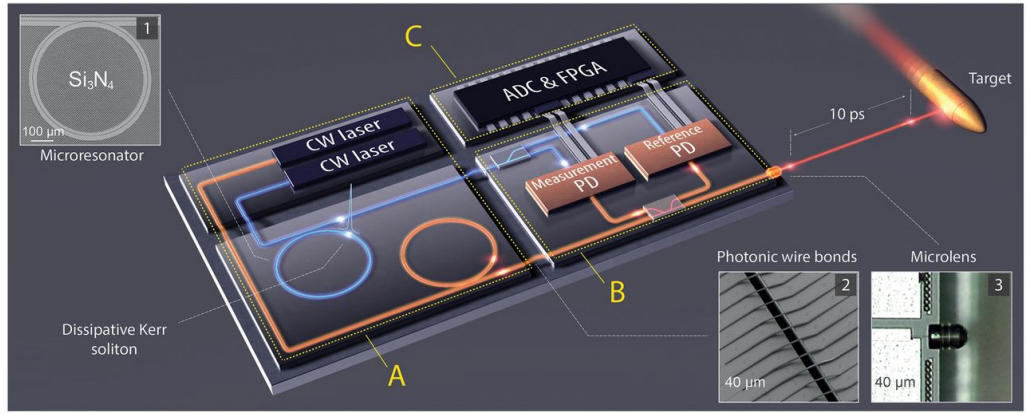

Fig. 26. Schematic of the ultrafast coherent laser ranging engine using dual-DKS microcomb. Reproduced with permission from Ref. [172].

Finally, quantum microcomb has attracted significant attention as the platform enables compact, controllable, manufacturable, and scalable complex quantum systems [[177], [178], [179], [180], [181], [182]]. Generations of frequency-multiplexed heralded photons and multiphoton, high-dimensional and hyper-entangled states have all been demonstrated recently. Aiming for scalable and practical quantum communication and computation, generation of multiphoton time-bin–entangled qubits [178] and control of two-photon high-dimensional frequency-entangled state [179,180] have both been demonstrated. Deterministic two-mode-squeezed quantum microcomb with 40 continuous-variable quantum modes also has been demonstrated at telecommunication wavelengths [181]. Furthermore, by using a wavelength division multiplexed microcomb, a parallel GHz quantum key distribution system with an ultrahigh secure key rate of more than 200 kbps at a distance of 25 km has been demonstrated [182]. With the continual efforts to understand and control new phenomena including spatiotemporal coupling [66,183], azimuthal cavity modulation [184], and coexisting Kerr and quadratic nonlinearities (see Section 4), the microcomb community will keep enhancing the performances of the photonic frequency microcomb and expanding the already remarkable microcomb application space.

**Funding.** National Science Foundation (ECCS 2048202, OMA 2016244) and Office of Naval Research (N00014-22-1-2224).

# Appendix

Table 6. Symbol definition in Section 3.

| Symbol | Definition | Symbol | Definition |
|---|---|---|---|
| $A$ | field envelope at $\omega_0$ | $\mathcal{F}^{-1}[\cdot]$ | inverse Fourier transformation |
| $B$ | field envelope at $2\omega_0$ | $\otimes$ | convolution operation |
| $A_{in}$ ($B_{in}$) | external cw driven field | $\Omega$ | angular frequency with respect to the field $A$ |
| $\alpha_{c1,2}$ | propagation loss per unit length, 1 for $A$, 2 for $B$ | $\alpha_{1,2}$ | total linear cavity loss |
| $\Delta k$ | wave-vector mismatch | $\xi = \Delta k\, L$ | wave-vector mismatch parameter |
| $\Delta k'$ | GVM | $d_1 = \Delta k'\, L$ | temporal walk-off |
| $k''_{1,2}$ | GVD | $d_2 = k''_2 L/2$ | group delay dispersion (GDD) for field $B$ |
| $\kappa = \sqrt{2}\omega_0 d_{eff}/\sqrt{A_{eff} c^3 n_1^2 n_2 \varepsilon_0}$ | normalized second-order nonlinear coupling coefficient | $t$ | slow time |
| $d_{eff}$ | effective second-order nonlinear coefficient | $t_R$ | roundtrip time |
| $A_{eff}$ | effective mode area | $\tau$ | fast time |
| $c$ | speed of light | $L$ | nonlinear medium length |
| $\varepsilon_0$ | vacuum permittivity | $\theta_{1,2}$ | coupling coefficients |
| $n_{1,2}$ | refractive index | $\Delta_2 = \delta_2/\alpha_2$ | normalized phase detuning for field $B$ |
| $\delta_{1,2}$ | resonance phase detunings | $D_2 = d_2/\alpha_2$ | normalized for field $B$ |
| | | $D_1 = d_1/\alpha_2$ | normalized temporal walk-off |